\documentclass[12pt]{article}

\usepackage{geometry, graphicx, subfig, float, hyperref, cite}
\usepackage{amsmath, amsfonts, amsthm, amssymb}
\numberwithin{equation}{section}

\begin{document}

\begin{titlepage}

\begin{center}
\vspace{1.0cm}
\Large{\textbf{Conformal blocks for highly disparate scaling dimensions}}\\
\vspace{0.8cm}
\small{\textbf{Connor Behan}}\\
\vspace{0.5cm}
\textit{Department of Physics and Astronomy, University of British Columbia, 6224 Agricultural Road, Vancouver BC, V6T 1Z1 Canada}
\end{center}

\vspace{1.0cm}
\begin{abstract}
We show that conformal blocks simplify greatly when there is a large difference between two of the scaling dimensions for external operators. In particular the spacetime dimension only appears in an overall constant which we determine via recurrence relations. Connections to the conformal bootstrap program and the AdS / CFT correspondence are also discussed.
\end{abstract}

\end{titlepage}

\tableofcontents

\section{Introduction}
Non-perturbative constraints on conformal field theories (CFTs) are useful for many areas of theoretical physics. An elementary fact about CFTs in arbitrary dimension is that the spacetime forms of the two and three point functions are fixed for all primary operators. Even though the four point function is not fixed by conformal symmetry, it is nevertheless constrained by crossing symmetry and the operator product expansion. These constraints led to a proposal by Polyakov in 1974 \cite{polyakov74} called the conformal bootstrap program. Starting with \cite{rattazzi08}, many works in recent years \cite{rychkov09, heemskerk09, caracciolo10, poland10, rattazzi10a, rattazzi10b, vichi11, poland11, showk12a, liendo12, showk12b, fitzpatrick12b, komargodski12, beem13, kos13, showk13, alday13, gaiotto13} have used the bootstrap approach to find general results about the space of possible CFTs. Special functions called conformal blocks have been an essential ingredient in all of these studies. A pedagogical discussion of them can be found in \cite{costa11}.

The conformal block for a primary operator $\mathcal{O}$, is the contribution of $\mathcal{O}$ and its descendants to the correlation function of four scalar primaries. We will let $\phi_i$ denote a scalar primary with dimension $\Delta_i$. Writing $x_{ij} \equiv x_i - x_j$ and $\Delta_{ij} \equiv \Delta_i - \Delta_j$, such a four point function is
\begin{equation}
\left < \phi_1(x_1) \phi_2(x_2) \phi_3(x_3) \phi_4(x_4) \right > = \left ( \frac{|x_{24}|}{|x_{14}|} \right )^{\Delta_{12}} \left ( \frac{|x_{14}|}{|x_{13}|} \right )^{\Delta_{34}} \frac{G(u, v)}{|x_{12}|^{\Delta_1 + \Delta_2}|x_{34}|^{\Delta_3 + \Delta_4}} \; , \label{4pt}
\end{equation}
where $G(u, v)$ is a function of the conformal invariants:
\begin{equation}
u = \frac{x_{12}^2 x_{34}^2}{x_{13}^2 x_{24}^2} \; , \; v = \frac{x_{14}^2 x_{23}^2}{x_{13}^2 x_{24}^2} \; . \label{uv-def}
\end{equation}
The fundamental tool for manipulating correlation functions is the operator product expansion (OPE):
\begin{equation}
\phi_1(x_1) \phi_2(x_2) = \sum_{\mathcal{O}} \lambda_{12 \mathcal{O}} C \left ( x_{12}, \frac{\partial}{\partial x_2} \right )^{\mu_1 \dots \mu_{\ell}} \mathcal{O}_{\mu_1 \dots \mu_{\ell}} (x_2) \; . \label{ope}
\end{equation}
Here, $C(x, \partial)$ is a differential operator which becomes $\frac{x^{\mu_1} \dots x^{\mu_{\ell}}}{|x|^{\Delta_1 + \Delta_2 - \Delta + \ell}}$ in the limit of small $x$. Taking the OPE in the $12$ and $34$ channels, (\ref{4pt}) can be rewritten as a linear combination:
\begin{equation}
\left < \phi_1(x_1) \phi_2(x_2) \phi_3(x_3) \phi_4(x_4) \right > = \left ( \frac{|x_{24}|}{|x_{14}|} \right )^{\Delta_{12}} \left ( \frac{|x_{14}|}{|x_{13}|} \right )^{\Delta_{34}} \sum_{\mathcal{O}} \lambda_{12 \mathcal{O}} \lambda_{34 \mathcal{O}} \frac{G_{\mathcal{O}}(u, v)}{|x_{12}|^{\Delta_1 + \Delta_2}|x_{34}|^{\Delta_3 + \Delta_4}} \; . \label{4pt-expanded}
\end{equation}
In the usual terminology, (\ref{4pt-expanded}) expresses $G(u, v)$ as a sum of \textit{conformal blocks} and it expresses the four point function as a sum of \textit{conformal partial waves}. The conformal block $G_{\mathcal{O}}(u, v)$ depends on the spin $\ell$ and scaling dimension $\Delta$ of the exchanged primary $\mathcal{O}$. It also depends on $\Delta_{12}$ and $\Delta_{34}$, the dimension differences of the external operators. Considering two, four and six spacetime dimensions, Dolan and Osborn found exact solutions \cite{dolan01, dolan04} for the blocks in terms of hypergeometric functions. In an odd (or fractional) number of dimensions, the conformal bootstrap must be carried out using one of the following more limited results:
\begin{enumerate}
\item Expressions in terms of Jack or Gegenbauer polynomials \cite{dolan04, hogervorst13a}.
\item Expressions in terms of Mellin-Barnes integrals \cite{mack09}.
\item A double power series for $\ell = 0$ together with recurrence relations \cite{dolan01, dolan11}.
\item The restriction to $(u - v + 1)^2 = 4u$ \cite{showk12a, hogervorst13b}.
\item The large $d$ limit \cite{fitzpatrick13}.
\end{enumerate}
In this note, we propose adding another special case to this list: the limit where $-\Delta_{12}$ is large.

The Casimir differential equation, which Dolan and Osborn used in their exact solutions, will be important for this derivation as well. In section 2, we review this equation and explain why an even $d$ makes it easier to solve than an odd $d$. In section 3, we make an ansatz for the conformal blocks with a highly disparate pair of scaling dimensions and demonstrate that it agrees with the solutions in even spacetime dimension. It is remarkably simple to show that the ansatz is correct for other spacetime dimensions as well up to an overall constant. Determining this constant is the focus of section 4.

Throughout these calculations, we may only consider large scaling dimension differences if at least one of the scaling dimensions themselves is large. Operators with large scaling dimensions lead to OPE coefficients that are exponentially small. Therefore, while $\Delta_2 \gg \Delta_1$ clearly leads to simpler expressions, it is not obvious that this case is interesting. Before we conclude, section 5 is used to argue that it is at least not obvious that this case is uninteresting.

\section{Main background}
An important result \cite{dolan04} reviewed in \cite{rattazzi08} is that the conformal blocks are eigenfunctions of a second order differential operator that is symmetric in two variables. Although not the first derivation of the exact conformal blocks \cite{dolan01}, this more intuitive treatment continues to be useful in understanding their properties \cite{dolan11, hogervorst13b, fitzpatrick13}. We will work in a Euclidean signature so that the conformal group is $SO(d + 1, 1)$ generated by four types of transformations: dilations $D$, translations $P_{\mu}$, special conformal transformations $K_{\mu}$ and rotations $M_{\mu\nu}$.

\subsection{The Casimir differential equation}
In a rotation group, the quadratic Casimir element is obtained by contracting the generators with themselves. In the conformal group, this Casimir is slightly different:
\begin{equation}
C_2 = \frac{1}{2} M_{\mu\nu} M^{\mu\nu} - D^2 - \frac{1}{2} \left ( P_{\mu}K^{\mu} + K_{\mu}P^{\mu} \right ) \; . \label{casimir}
\end{equation}
This can either be found by brute force, or by realizing that the conformal group is the Lorentz group of a higher dimensional space. Primary operators transform under the generators according to
\begin{eqnarray}
\left [ D, \mathcal{O} \right ] &=& i \left ( x^{\mu} \partial_{\mu} + \Delta \right ) \mathcal{O} \nonumber \\
\left [ P_{\mu}, \mathcal{O} \right ] &=& i \partial_{\mu} \mathcal{O} \nonumber \\
\left [ K_{\mu}, \mathcal{O} \right ] &=& i \left ( x^2 \partial_{\mu} - 2x_{\mu} x^{\nu} \partial_{\nu} - 2x_{\mu}\Delta + 2x^{\nu}M^{\ell}_{\mu\nu} \right ) \mathcal{O} \nonumber \\
\left [ M_{\mu\nu}, \mathcal{O} \right ] &=& i \left ( x_{\mu} \partial_{\nu} - x_{\nu} \partial_{\mu} + M^{\ell}_{\mu\nu} \right ) \mathcal{O} \nonumber
\end{eqnarray}
where $M^{\ell}_{\mu\nu}$ is the rotation operator in the spin $\ell$ representation \cite{minwalla97}. Using these transformations and antisymmetry of $M^{\ell}_{\mu\nu}$, it can be shown that primary operators satisfy the eigenvalue equation
\begin{eqnarray}
C_2 \mathcal{O} &=& -2 \Lambda_d \mathcal{O} \nonumber \\
2 \Lambda_d &=& \Delta ( \Delta - d ) + \ell ( \ell + d - 2 ) \; . \label{primary-eigenvalue}
\end{eqnarray}
Since the Casimir for a Lie algebra is part of the center, (\ref{primary-eigenvalue}) holds for all descendants of $\mathcal{O}$ as well. Therefore, the conformal partial wave for $\mathcal{O}$ in $\left < \phi_1(x_1) \phi_2(x_2) \phi_3(x_3) \phi_4(x_4) \right >$ should have the same eigenvalue under $C_2$. When we apply $C_2$ to the partial wave in (\ref{4pt-expanded}), we need two copies of each generator because $\mathcal{O}$ comes from the OPE of $\phi_1(x_1)$ and $\phi_2(x_2)$. Instead of writing $\left ( M^1_{\mu\nu} + M^2_{\mu\nu} \right ) \left ( M_1^{\mu\nu} + M_2^{\mu\nu} \right )$ \textit{etc}, it is much easier to write
\begin{equation}
C_2 = \frac{1}{2} \left ( L^1_{AB} + L^2_{AB} \right ) \left ( L^1_{AB} + L^2_{AB} \right ) \nonumber
\end{equation}
where $L^i_{AB} = P^i_A \frac{\partial}{\partial P^i_B} - P^i_B \frac{\partial}{\partial P^i_A}$ acts on the $d + 2$ dimensional null rays $P^i$. Lifting (\ref{4pt-expanded}) into this space, the object we must differentiate is
\begin{equation}
W_{\mathcal{O}}(P_1, P_2, P_3, P_4) = \left ( \frac{P_2 \cdot P_4}{P_1 \cdot P_4} \right )^{\frac{1}{2} \Delta_{12}} \left ( \frac{P_1 \cdot P_4}{P_1 \cdot P_3} \right )^{\frac{1}{2} \Delta_{34}} \frac{G_{\mathcal{O}}(u, v)}{(P_1 \cdot P_2)^{\frac{1}{2}(\Delta_1 + \Delta_2)} (P_3 \cdot P_4)^{\frac{1}{2}(\Delta_3 + \Delta_4)}} \label{4pt-embedded}
\end{equation}
with $u$ and $v$ defined by changing $x_{ij}^2$ to $P_i \cdot P_j$. The computations reveal the following equation for the conformal blocks:
\begin{align}
& \left [ -2v \left ( (1 - v)^2 - u(1 + v) \right ) \frac{\partial^2}{\partial v^2} + 4 uv (1 + u - v) \frac{\partial^2}{\partial u \partial v} - 2u^2 (1 - u + v) \frac{\partial^2}{\partial u^2} \right ] G_{\mathcal{O}} \nonumber \\
& + \left [ -2 \left ( (1 - v)^2 - u(1 + v) \right ) \frac{\partial}{\partial v} - 2u (1 - d - u + v) \frac{\partial}{\partial u} \right ] G_{\mathcal{O}} \nonumber \\
& - \left ( \Delta_{12} - \Delta_{34} \right ) \left [ (1 + u - v) \left ( u \frac{\partial}{\partial u} + v \frac{\partial}{\partial v} \right ) - (1 - u - v) \frac{\partial}{\partial v} \right ] G_{\mathcal{O}} \nonumber \\
& - \frac{1}{2} \Delta_{12} \Delta_{34} (u - v + 1) G_{\mathcal{O}} = -2\Lambda_d G_{\mathcal{O}} \; . \label{big-uv-equation}
\end{align}
After deriving (\ref{big-uv-equation}), Dolan and Osborn made it symmetric with the following change of variables \cite{dolan04}:
\begin{equation}
u = xz \; , \; v = (1-x)(1-z) \; . \label{xz-def}
\end{equation}
This leads to the equation $D_d G_{\mathcal{O}} = \Lambda_d G_{\mathcal{O}}$, where
\begin{eqnarray}
D_d &=& x^2 (1 - x) \frac{\partial^2}{\partial x^2} + x(c - (a + b + 1)x) \frac{\partial}{\partial x} - abx + (x \leftrightarrow z) \nonumber \\
&& + (d - 2) \frac{xz}{x-z} \left ( (1 - x) \frac{\partial}{\partial x} - (1 - z) \frac{\partial}{\partial z} \right ) \label{differential-operator}
\end{eqnarray}
for $a = -\frac{1}{2} \Delta_{12}$, $b = \frac{1}{2} \Delta_{34}$ and $c = 0$.

\subsection{Known solutions}
An important eigenvalue equation to use when trying to solve this is
\begin{eqnarray}
&& \left [ x^2 (1 - x) \frac{\textup{d}^2}{\textup{d} x^2} + x(c - (a + b + 1)x) \frac{\textup{d}}{\textup{d} x} - abx \right ] x^{\lambda} {}_2F_1 (\lambda + a, \lambda + b; 2\lambda + c; x) \nonumber \\
&=& \lambda (\lambda + c - 1) x^{\lambda} {}_2F_1 (\lambda + a, \lambda + b; 2\lambda + c; x) \; . \label{hypergeometric-operator}
\end{eqnarray}
When $d = 2$, this can be used immediately to say that
\begin{equation}
G_{\mathcal{O}}(x, z) = x^{\lambda_1} z^{\lambda_2} {}_2F_1 (\lambda_1 + a, \lambda_1 + b; 2\lambda_1 + c; x) {}_2F_1 (\lambda_2 + a, \lambda_2 + b; 2\lambda_2 + c; z) + (x \leftrightarrow z) \label{2d-solution}
\end{equation}
is an eigenfunction of $D_d$ with eigenvalue $\lambda_1 (\lambda_1 - 1) + \lambda_2 (\lambda_2 + 1 - d)$. This eigenvalue is precisely $\Lambda_d$ if
\begin{equation}
\lambda_1 = \frac{1}{2} (\Delta + \ell) \; , \; \lambda_2 = \frac{1}{2} (\Delta - \ell) \; . \label{lambda-def}
\end{equation}

For higher dimensions (\ref{differential-operator}) has coupling between $x$ and $z$. This case is handled by
introducing the functions
\begin{equation}
F_{p,q}^{\pm,r}(x, z) = x^{p + r} z^{q + r} {}_2F_1(p + a, p + b; 2p + c; x) {}_2F_1(q + a, q + b; 2q + c; z) \pm (x \leftrightarrow z) \; . \label{helpful-functions}
\end{equation}
The known solutions are of the form
\begin{equation}
G_{\mathcal{O}}(x, z) = \sum_{p, q} a_{p, q} \frac{F_{p,q}^{\pm,r}(x, z)}{(x - z)^s} \label{form1}
\end{equation}
where we choose the positive sign for even $s$ and the negative sign for odd $s$. To fix the unspecified parameters in (\ref{form1}), the first step is to make $D_d G_{\mathcal{O}}$ as simple as possible by choosing $s$ appropriately. A straightforward calculation tells us that
\begin{eqnarray}
(x - z)^s D_d \frac{F}{(x - z)^s} &=& D_2 F - sc F - s(a + b + 1)(x + z) F - s(d - 2)(x + z - 1) F \nonumber \\
&& + s(s - d + 3) \frac{x^2 (1 - x) + z^2 (1 - z)}{(x - z)^2} F - 2s \left ( x(1 - x) \frac{\partial F}{\partial x} + z(1 - z) \frac{\partial F}{\partial z} \right ) \nonumber \\
&& + (d - 2 - 2s) \frac{xz}{x-z} \left ( (1 - x) \frac{\partial F}{\partial x} - (1 - z) \frac{\partial F}{\partial z} \right ) \label{before-s}
\end{eqnarray}
The choice that allows us to absorb most of the extra terms into $D_2$ is $s = d - 3$. This gives us the equation:
\begin{eqnarray}
D_d \frac{F}{(x - z)^{d - 3}} &=& \frac{1}{(x - z)^{d - 3}} \left [ \biggl. D_2 \biggl |_{\substack{a \mapsto a - d + 3, b \mapsto b - d + 3 \\ c \mapsto c - 2d + 6}} + (d - 3)(d - 2 - c) \right. \nonumber \\
&& \left. - (d - 4) \frac{xz}{x - z} \left ( (1 - x) \frac{\partial}{\partial x} - (1 - z) \frac{\partial}{\partial z} \right ) \right ] F \; . \label{after-s}
\end{eqnarray}
In order to benefit from the simplified form of (\ref{after-s}), we want our $F_{p,q}^{\pm,r}$ to be eigenfunctions of the shifted $D_2$ operator. This forces us to set $r = d - 3$ as well. Inserting
\begin{equation}
G_{\mathcal{O}}(x, z) = \sum_{p, q} a_{p, q} \frac{F_{p,q}^{\pm,d - 3}(x, z)}{(x - z)^{d - 3}} \label{form2}
\end{equation}
into $D_d G_{\mathcal{O}} = \Lambda_d G_{\mathcal{O}}$, we arrive at
\begin{eqnarray}
&& \sum_{p, q} a_{p, q} [ p(p + c - 1) + q(q + c - 1) - (d - 3)(d - 2 - c) - \Lambda_d ] \left ( \frac{1}{x} - \frac{1}{z} \right ) F_{p,q}^{\pm,d - 3}(x, z) \nonumber \\
&=& -(d - 4) \sum_{p, q} a_{p, q} \left ( (1 - x) \frac{\partial}{\partial x} - (1 - z) \frac{\partial}{\partial z} \right ) F_{p,q}^{\pm,d - 3}(x, z) \; . \label{helpful-equation}
\end{eqnarray}
There are recurrence relations satisfied by the $F_{p,q}^{\pm,r}$ that we can use on either side of (\ref{helpful-equation}). Defining
\begin{eqnarray}
A_{p, q} &=& 2 (p - q)(p + q + c - 1) \frac{(2a - c)(2b - c)}{(2p + c)(2p + c - 2)(2q + c)(2q + c - 2)} \nonumber \\
B_p &=& \frac{(p + a)(p + b)(p + c - a)(p + c - b)}{(2p + c - 1)(2p + c)^2(2p + c + 1)} \; , \label{ab-def}
\end{eqnarray}
they are:
\begin{align}
& \left ( \frac{1}{x} - \frac{1}{z} \right ) F_{p,q}^{\pm,r}(x, z) = F_{p-1,q}^{\mp,r}(x, z) - F_{p,q-1}^{\mp,r}(x, z) + A_{p, q} F_{p,q}^{\mp,r}(x, z) + B_p F_{p+1,q}^{\mp,r}(x, z) - B_q F_{p,q+1}^{\mp,r}(x, z) \nonumber \\
& \left ( (1 - x) \frac{\partial}{\partial x} - (1 - z) \frac{\partial}{\partial z} \right ) F_{p,q}^{\pm,r}(x, z) = (p + r) F_{p-1,q}^{\mp,r}(x, z) - (q + r) F_{p,q-1}^{\mp,r}(x, z) \nonumber \\
& - \frac{c - 2r - 2}{2} A_{p, q} F_{p,q}^{\mp,r}(x, z) - (p + c - 1 - r) B_p F_{p+1,q}^{\mp,r}(x, z) + (q + c - 1 - r) F_{p,q+1}^{\mp,r}(x, z) \; . \label{recursion1}
\end{align}
If we substitute (\ref{recursion1}) into (\ref{helpful-equation}) and reindex the sum, we find a recurrence relation among the coefficients:
\begin{align}
& 0 = a_{p + 1, q} [(p + 1)(p + c) + (d - 4)(p + d - 2) + q(q + c - 1) - (d - 3)(d - 2 - c) - \Lambda_d] \nonumber \\
& - a_{p, q + 1} [p(p + c - 1) + (q + 1)(q + c) + (d - 4)(q + d - 2) - (d - 3)(d - 2 - c) - \Lambda_d] \nonumber \\
& + a_{p, q} A_{p, q} \left [ p(p + c - 1) + q(q + c - 1) - (d - 4) \frac{c - 2d - 4}{2} - (d - 3)(d - 2 - c) - \Lambda_d \right ] \nonumber \\
& + a_{p - 1, q} B_{p - 1} [(p - 1)(p + c - 2) - (d - 4)(p + c - d + 1) + q(q + c - 1) - (d - 3)(d - 2 - c) - \Lambda_d] \nonumber \\
& - a_{p, q - 1} B_{q - 1} [p(p + c - 1) + (q - 1)(q + c - 2) - (d - 4)(q + c - d + 1) - (d - 3)(d - 2 - c) - \Lambda_d] \; . \label{recursion2}
\end{align}
Our ability to solve this for only finitely many $a_{p, q} \neq 0$ depends on $d$ being even.
\begin{figure}[h]
\centering
\includegraphics[scale=0.3]{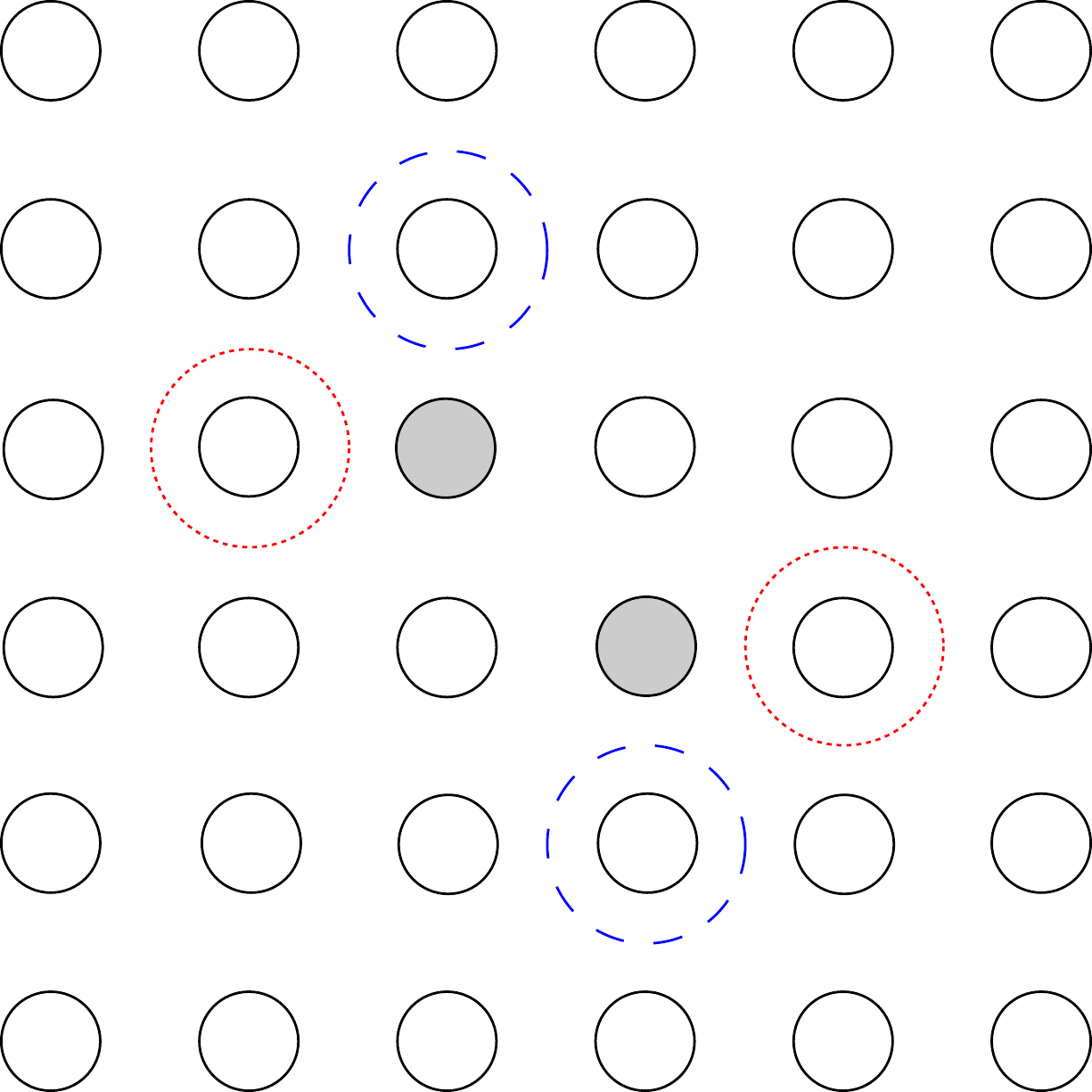}
\caption{This simple shape has exactly one boundary point of each type. The left and right boundary points as we are defining them are circled with dotted red lines. The top and bottom boundary points are circled with dashed blue lines.}
\label{lattice-example}
\end{figure}
To help explain this, we can draw a lattice with integer spacing where $a_{p, q} \neq 0$ if and only if the circle at $(p, q)$ is shaded. Every shape with finitely many shaded points must have a left boundary point. This is an unshaded circle where all of its nearest neighbours are unshaded except the one to the left. Similarly, it must also have a right boundary point. The $p$-coordinates of these points will differ by at least two integer units $m$. If the shape is reasonably symmetric, their $q$-coordinates can be chosen to be the same. Therefore, we are dealing with $(p_*, q_*)$ and $(p_* + m, q_*)$. The left boundary point requires the coefficient on $a_{p + 1, q}$ to vanish when $(p, q) = (p_*, q_*)$. The right boundary point requires the coefficient on $a_{p - 1, q}$ to vanish when $(p, q) = (p_* + m, q_*)$. Setting these coefficients equal to each other, we have:
\begin{eqnarray}
p_*^2 + (d - 3)p_* + (d - 4)(d - 2) &=& (p_* + m)^2 - (d - 1)(p_* + m) + (d - 4)(d - 1) + 2 \nonumber \\
2(m - d + 2)p_* &=& (m - d + 2)(1 - m) \; . \nonumber
\end{eqnarray}
If $m \neq d - 2$, our left and right boundary points are $(\frac{1 - m}{2}, q_*)$ and $(\frac{1 + m}{2}, q_*)$ respectively. The same analysis tells us that the top and bottom boundary points are $(p^*, \frac{1 + n}{2})$ and $(p^*, \frac{1 - n}{2})$ respectively. These expressions are not compatible for general $\lambda_1$ and $\lambda_2$. In order for the aforementioned coefficients to be zero, $q_*$ depends on $\lambda_1$ and $\lambda_2$ in such a way that it is almost surely not an integer number of units away from the half integers $\frac{1 \pm n}{2}$. Similarly, the value that $p^*$ must have is almost surely not an integer away from $\frac{1 \pm m}{2}$. We should instead consider $m = n = d - 2$. This property holds for the known solutions defined below \cite{dolan04}.
\begin{align}
\begin{tabular}{|c|l|}
\hline
$d = 4$ & $\;\;\; a_{\lambda_1, \lambda_2 - 1} = 1$ \\
\hline
& $\;\;\; a_{\lambda_1, \lambda_2 - 3} = 1$ \\
& $a_{\lambda_1 - 1, \lambda_2 - 2} = -\frac{\lambda_1 - \lambda_2 + 3}{\lambda_1 - \lambda_2 + 1}$ \\
$d = 6$ & $\;\;\; a_{\lambda_1, \lambda_2 - 2} = -\frac{(\lambda_1 + \lambda_2 + c - 4)(\lambda_1 - \lambda_2 + 3)}{(\lambda_1 + \lambda_2 + c - 3)(\lambda_1 - \lambda_2 + 2)} A_{\lambda_1, \lambda_2 - 2}$ \\
& $\;\;\; a_{\lambda_1, \lambda_2 - 1} = \frac{(\lambda_1 + \lambda_2 + c - 4)(\lambda_1 - \lambda_2 + 3)}{(\lambda_1 + \lambda_2 + c - 2)(\lambda_1 + \lambda_2 + 1)} B_{\lambda_2 - 2}$ \\
& $a_{\lambda_1 + 1, \lambda_2 - 2} = -\frac{\lambda_1 + \lambda_2 + c - 4}{\lambda_1 + \lambda_2 + c - 2} B_{\lambda_1}$ \\
\hline
\end{tabular}
\label{exact-solutions}
\end{align}
\begin{figure}[h]
\centering
\subfloat[][$d = 4$]{\includegraphics[scale=0.3]{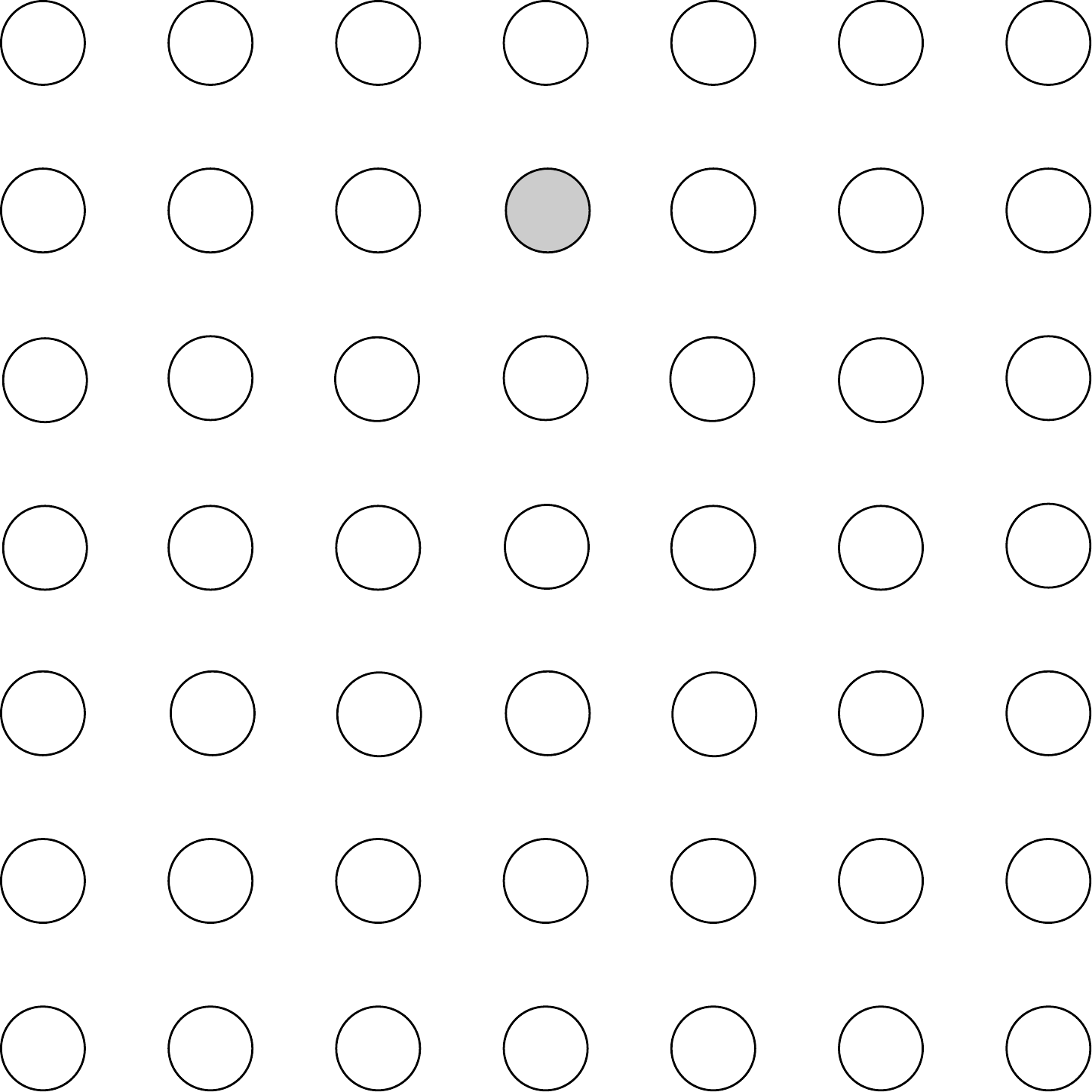}} \qquad
\subfloat[][$d = 6$]{\includegraphics[scale=0.3]{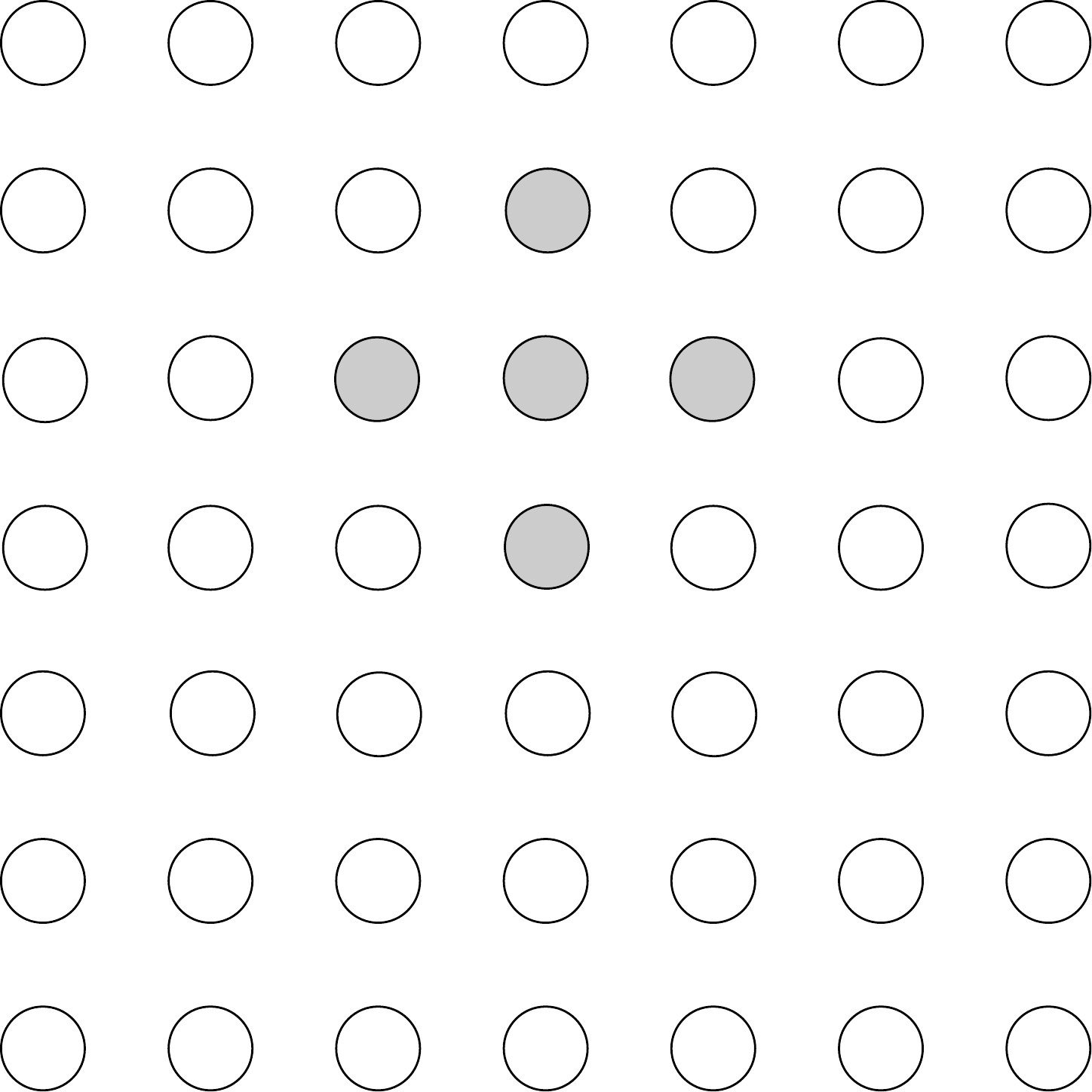}} \qquad
\subfloat[][$d = 8$?]{\includegraphics[scale=0.3]{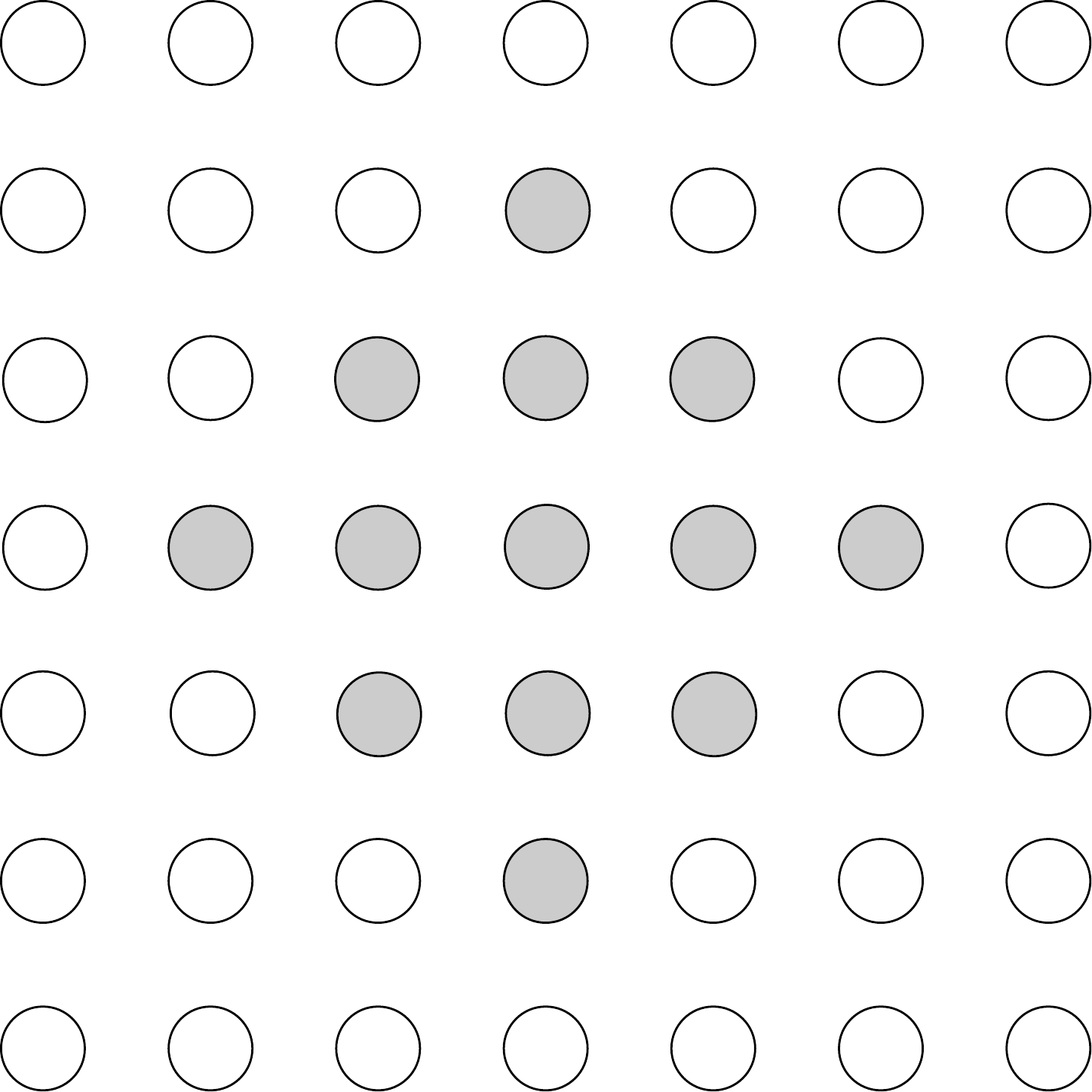}}
\caption{The solution in four dimensions corresponds to the point at $(\lambda_1, \lambda_2 - 1)$. In six dimensions, this becomes 5 points centred at $(\lambda_1, \lambda_2 - 2)$. If one were to solve for conformal blocks in eight dimensions, this would most likely involve the 13 points centred at $(\lambda_1, \lambda_2 - 3)$.}
\label{lattice-diamonds}
\end{figure}
\begin{figure}[h]
\centering
\includegraphics[scale=0.3]{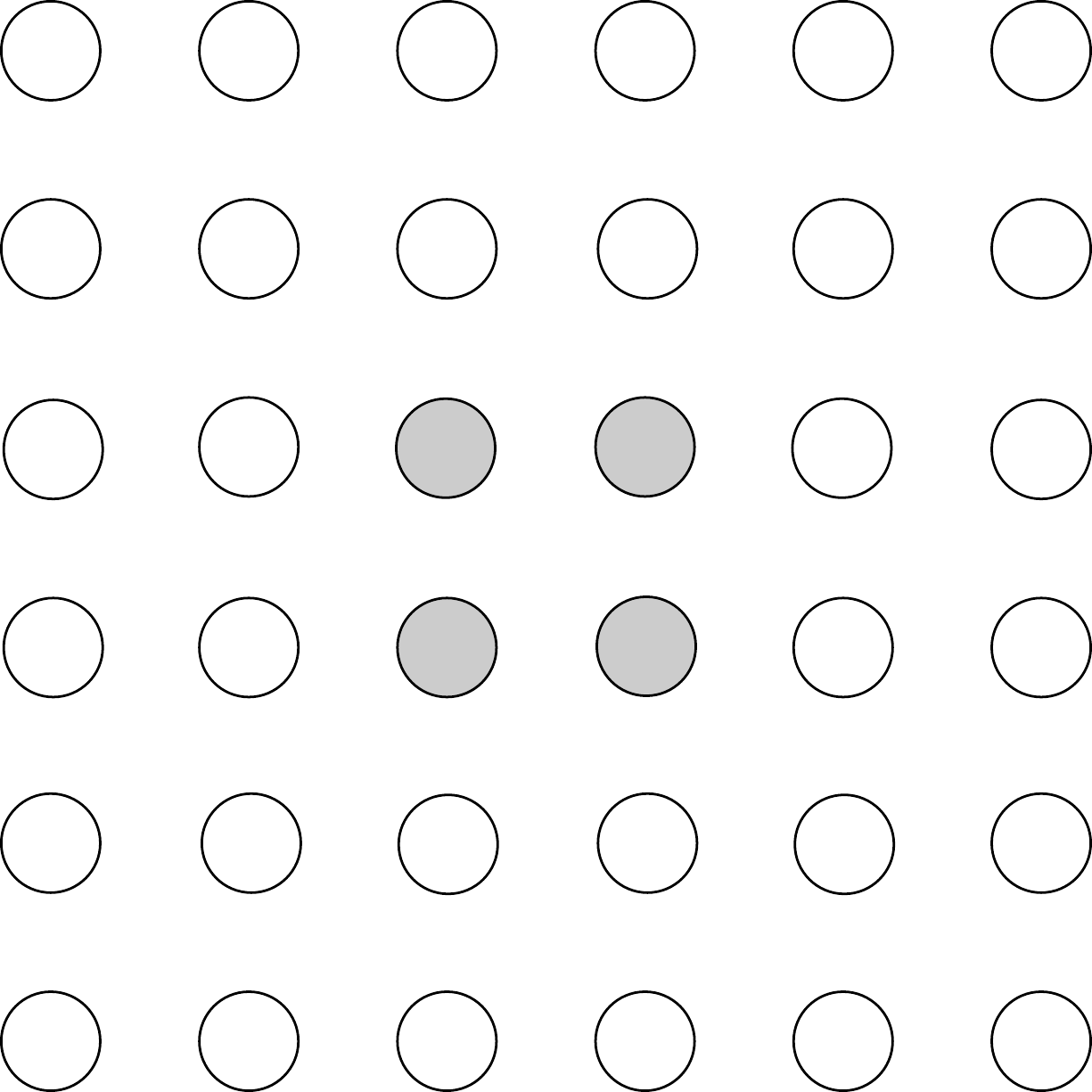}
\caption{Each boundary point for this square has a boundary point across from it that is three units away. However, this is not a sufficient shape for the coefficients in five dimensions because some of the boundary points are adjacent.}
\label{lattice-square}
\end{figure}
In four dimensions, the width and height equal to $2$ leave only a single shaded circle. In six dimensions, the width and height equal to $4$ leave the diamond of five shaded circles shown in Figure \ref{lattice-diamonds}. We run into a problem when we try to let the distance between boundary points be an odd integer. We are forced to draw a square instead of a diamond as the most symmetric shape having this property. From (\ref{recursion2}), it is clear that if a certain coefficient vanishes at $(p, q)$, it will not vanish at $(p \pm 1, q)$ or $(p, q \pm 1)$. Therefore the recurrence relation is not compatible with a square because it does not allow boundary points to be nearest neighbours.

\section{The highly disparate limit}
To approximate conformal blocks for large parameters, it will be useful to first do this for the hypergeometric function. Defining the Pochhammer symbol by $(x)_n = \frac{\Gamma(x + n)}{\Gamma(x)}$, the Gauss hypergeometric function is
\begin{equation}
{}_2F_1(a, b; c; x) = \sum_{n = 0}^{\infty} \frac{(a)_n (b)_n}{(c)_n} \frac{x^n}{n!} \label{gauss}
\end{equation}
in the unit disk. We can see the asymptotic behaviour more easily from the Euler integral which holds for $\Re{c} > \Re{b} > 0$:
\begin{equation}
B(b, c - b) {}_2F_1(a, b; c; x) = \int_0^1 t^{b - 1}(1 - t)^{c - b - 1}(1 - tx)^{-a} \textup{d}t \; . \label{euler}
\end{equation}
To expand this integral for large $a$, one writes $(1 - tx)^{-a} = e^{-af(t)}$ where $f(t) = \log(1 - tx)$. Since $0 < x < 1$, the minimum of $f$ on the interval $[0, 1]$ is achieved at $t = 1$. The saddle point approximation therefore takes the form:
\begin{equation}
e^{-af(t)} \approx e^{-af(1) - a(t - 1)f^{\prime}(1)} = (1 - x)^{-a} e^{(t - 1) \frac{ax}{1 - x}} \; . \nonumber
\end{equation}
This is different from the usual expression $e^{-af(t)} \approx e^{-af(t_0) - a\frac{(t - t_0)^2}{2}f^{\prime\prime}(t_0)}$ because the function was minimized at an endpoint. The resulting integral substituting $s$ for $1 - t$ is:
\begin{eqnarray}
B(b, c - b) {}_2F_1(a, b; c; x) &\approx& (1 - x)^{-a} \int_0^1 e^{-s \frac{ax}{1 - x}} s^{c - b - 1} (1 - s)^{b - 1} \textup{d}s \nonumber \\
&\approx& (1 - x)^{-a} \int_0^1 e^{-s \frac{ax}{1 - x}} s^{c - b - 1} \textup{d}s \nonumber \\
&\approx& (1 - x)^{-a} \int_0^{\infty} e^{-s \frac{ax}{1 - x}} s^{c - b - 1} \textup{d}s \nonumber \\
&=& \Gamma(c - b) (1 - x)^{-a} \left ( \frac{1 - x}{ax} \right )^{c - b} \; . \label{2f1-approx1}
\end{eqnarray}
We neglected $(1 - u)^{b - 1}$ because $u = 0$ provides the dominant contribution to the integral. The upper limit can be changed from $1$ to $\infty$ if we are only going to leading order.

\subsection{Asymptotic eigenfunctions}
We will now find the large $a$ limit of the conformal blocks. To do this, we do not need to know a full solution to the Casimir differential equation, only the first few terms in its asymptotic series. This means that instead of $D_d G_{\mathcal{O}} = \Lambda_d G_{\mathcal{O}}$, we may consider the easier problem $D_d G_{\mathcal{O}} = \Lambda_d G_{\mathcal{O}} \left ( 1 + O \left ( \frac{1}{a} \right ) \right )$. Motivated by (\ref{2f1-approx1}), we expect
\begin{equation}
G^{(0)}_{\mathcal{O}}(x, z) = \frac{(1 - x)^{-a}(1 - z)^{-a}}{a^{\lambda_1 + \lambda_2 - 2b}} \left ( \frac{xz}{(1 - x)(1 - z)} \right )^b \label{result1}
\end{equation}
to be the leading behaviour. For any number of dimensions, the smallest power of $\frac{1}{a}$ in $D_d G^{(0)}_{\mathcal{O}}(x, z)$ is $\lambda_1 + \lambda_2 - 2b + 1$. Therefore, we need at least one more term to produce $\Lambda_d G^{(0)}_{\mathcal{O}}(x, z)$.\footnote{This is clear from the fact that (\ref{result1}) is symmetric in $\lambda_1$ and $\lambda_2$ but the desired eigenvalue is not.} Choosing
\begin{equation}
G^{(1)}_{\mathcal{O}}(x, z) = \frac{(1 - x)^{-a}(1 - z)^{-a}}{2a^{\lambda_1 + \lambda_2 - 2b + 1}} \left ( \frac{xz}{(1 - x)(1 - z)} \right )^b \left ( \gamma - \frac{1}{x} - \frac{1}{z} \right ) \left [ \Lambda_d - b(2b - d) \right ]\label{result2}
\end{equation}
for some $\gamma$, this can be made to work as follows:
% Very hackish.
\begin{align}
D_d \left ( G^{(0)}_{\mathcal{O}}(x, z) \right. &+ \left. G^{(1)}_{\mathcal{O}}(x, z) \right ) = \frac{z^b(1 - z)^{-a - b}}{a^{\lambda_1 + \lambda_2 - 2b}} \left [ x^2(1 - x) \frac{\partial}{\partial x} - x^2 (a + b + 1) + (d - 2) \frac{xz(1 - x)}{x - z} \right ] \nonumber \\
& \left [ \left ( bx^{b - 1}(1 - x)^{-a - b} + (a + b)x^b(1 - x)^{-a - b - 1} \right ) \left ( 1 + \frac{\Lambda_d - b(2b - d)}{2a} \left ( \gamma - \frac{1}{x} - \frac{1}{z} \right ) \right ) \right. \nonumber \\
& \left. + x^{b - 2}(1 - x)^{-a - b} \frac{\Lambda_d - b(2b - d)}{2a} \right ] - abx \frac{x^b(1 - x)^{-a - b}z^b(1 - z)^{-a - b}}{a^{\lambda_1 + \lambda_2 - 2b}} \nonumber \\
& \left ( 1 + \frac{\Lambda_d - b(2b - d)}{2a} \left ( \gamma - \frac{1}{x} - \frac{1}{z} \right ) \right ) + (x \leftrightarrow z) \nonumber \\
& \;\;\;\;\;\;\;\;\;\;\;\;\;\;\;\;\;\;\;\;\; = \frac{x^2 (1 - x)^{-a - b - 1} z^b(1 - z)^{-a - b}}{a^{\lambda_1 + \lambda_2 - 2b}} \nonumber \\ 
& \left [ \left ( b(b - 1)x^{b - 2}(1 - x)^2 + 2b(a + b)x^{b - 1}(1 - x) + (a + b)(a + b + 1)x^b \right ) \right. \nonumber \\
& \left ( 1 + \frac{\Lambda_d - b(2b - d)}{2a} \left ( \gamma - \frac{1}{x} - \frac{1}{z} \right ) \right ) + \left ( (b - 1)x^{b - 1}(1 - x)^2 +(a + b)x^b(1 - x) \right ) \nonumber \\
& \left. \frac{\Lambda_d - b(2b - d)}{ax^2} \right ] + \frac{z^b(1 - z)^{-a - b}}{a^{\lambda_1 + \lambda_2 - 2b}} \left [ -x^2 (a + b + 1) + (d - 2) \frac{xz(1 - x)}{x - z} \right ] \nonumber \\
& \left [ \left ( bx^{b - 1}(1 - x)^{-a - b} + (a + b)x^b(1 - x)^{-a - b - 1} \right ) \left ( 1 + \frac{\Lambda_d - b(2b - d)}{2a} \left ( \gamma - \frac{1}{x} - \frac{1}{z} \right ) \right ) \right. \nonumber \\
& \left. + x^{b - 2}(1 - x)^{-a - b} \frac{\Lambda_d - b(2b - d)}{2a} \right ] - abx \frac{x^b(1 - x)^{-a - b}z^b(1 - z)^{-a - b}}{a^{\lambda_1 + \lambda_2 - 2b}} \nonumber \\
& \left ( 1 + \frac{\Lambda_d - b(2b - d)}{2a} \left ( \gamma - \frac{1}{x} - \frac{1}{z} \right ) \right ) + (x \leftrightarrow z) \nonumber \\
& \;\;\;\;\;\;\;\;\;\;\;\;\;\;\;\;\;\;\;\;\; = \frac{x^b(1 - x)^{-a - b}z^b(1 - z)^{-a - b}}{a^{\lambda_1 + \lambda_2 - 2b}} \nonumber \\
& \left [ \frac{1}{2} \left ( \Lambda_d - b(2b - d) \right ) + b(b - 1) + b(d - 2) \frac{xz}{x - z} \frac{1 - x}{x} \right ] + O \left ( \frac{1}{a^{\lambda_1 + \lambda_2 - 2b + 1}} \right ) + (x \leftrightarrow z) \nonumber \\
& \;\;\;\;\;\;\;\;\;\;\;\;\;\;\;\;\;\;\;\;\; = \Lambda_d \frac{x^b(1 - x)^{-a - b}z^b(1 - z)^{-a - b}}{a^{\lambda_1 + \lambda_2 - 2b}} \left ( 1 + O \left ( \frac{1}{a} \right ) \right ) \nonumber \\
& \;\;\;\;\;\;\;\;\;\;\;\;\;\;\;\;\;\;\;\;\; = \Lambda_d G^{(0)}_{\mathcal{O}}(x, z) \left ( 1 + O \left ( \frac{1}{a} \right ) \right ) \nonumber \\
& \;\;\;\;\;\;\;\;\;\;\;\;\;\;\;\;\;\;\;\;\; = \Lambda_d \left ( G^{(0)}_{\mathcal{O}}(x, z) + G^{(1)}_{\mathcal{O}}(x, z) \right ) \left ( 1 + O \left ( \frac{1}{a} \right ) \right ) \; . \label{long-calculation}
\end{align}
The longest line in (\ref{long-calculation}) has terms of order $a^{2b - \lambda_1 - \lambda_2 + 2}$ but one can quickly see that their coefficients sum to zero. The terms of order $a^{2b - \lambda_1 - \lambda_2 + 1}$ can be seen to vanish almost as quickly; one has to use $(x \leftrightarrow z)$ to cancel an antisymmetric function. Since eigenvalue equations are linear, our asymptotic (\ref{result1}) can be multiplied by a constant depending on the dimension. Restating our main result in less specialized notation,
\begin{equation}
G_{\mathcal{O}}(u, v) \sim C^{(d)}_{\mathcal{O}} \frac{v^{\frac{1}{2}\Delta_{12}}}{\left | \frac{1}{2} \Delta_{12} \right |^{\Delta - \Delta_{34}}} \left ( \frac{u}{v} \right )^{\frac{1}{2} \Delta_{34}} \; , \; \Delta_{12} \rightarrow -\infty \; . \label{result3}
\end{equation}

\subsection{Agreement with exact blocks}
Even though we have already proven our main result (\ref{result3}), it is interesting to see how it is reproduced by the conformal blocks in even spacetime dimension. For the $d = 2$ solution (\ref{2d-solution}), one may expand the hypergeometric functions to first order to arrive at:
\begin{equation}
G_{\mathcal{O}}(x, z) \sim 2 \frac{\Gamma(2\lambda_1)\Gamma(2\lambda_2)}{\Gamma(\lambda_1 + b)\Gamma(\lambda_2 + b)} G^{(0)}_{\mathcal{O}}(x, z) \; . \label{result2d}
\end{equation}
The second asymptotic term in the two dimensional blocks can only match (\ref{result2}) if a single hypergeometric function has
\begin{equation}
{}_2F_1(a, b; c; x) \sim \frac{\Gamma(c)}{\Gamma(b)} (1 - x)^{-a} \left ( \frac{1 - x}{ax} \right )^{c - b} \left [ 1 + \frac{(c - b)(b - 1)}{a} \left ( \frac{\gamma}{2} - \frac{1}{x} \right ) \right ] \label{2f1-approx2}
\end{equation}
as its first correction beyond the saddle point. We will not need to know $\gamma$ for these calculations but we would see that $\gamma = 1$ if we followed the procedure in \cite{temme03}.\footnote{This is not the value we would get if we naively used the binomial theorem in (\ref{2f1-approx1}) and tried to integrate from $0$ to $\infty$ again.}
\begin{figure}[h]
\begin{tabular}{ccc}
\includegraphics[scale=0.2]{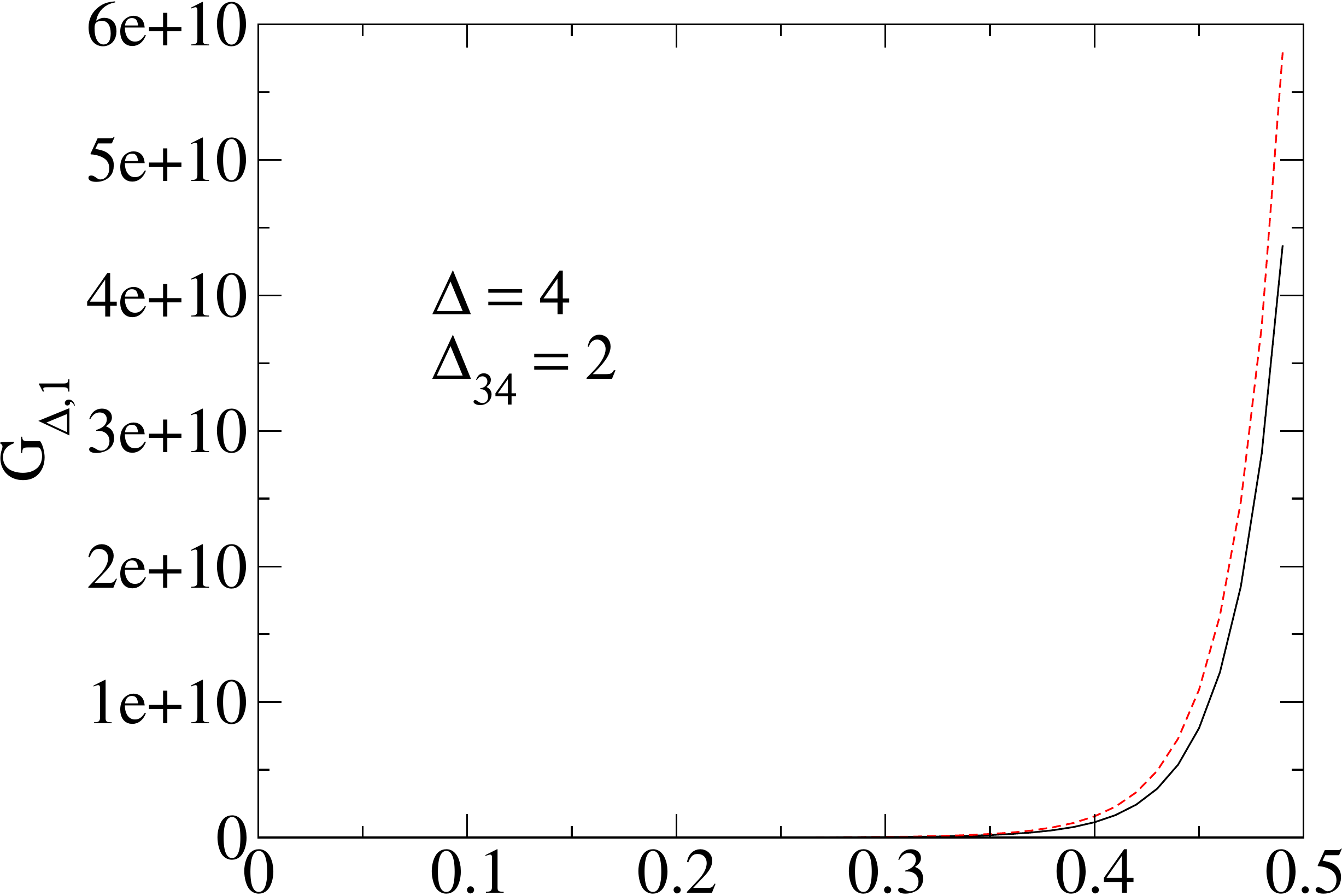} & \includegraphics[scale=0.2]{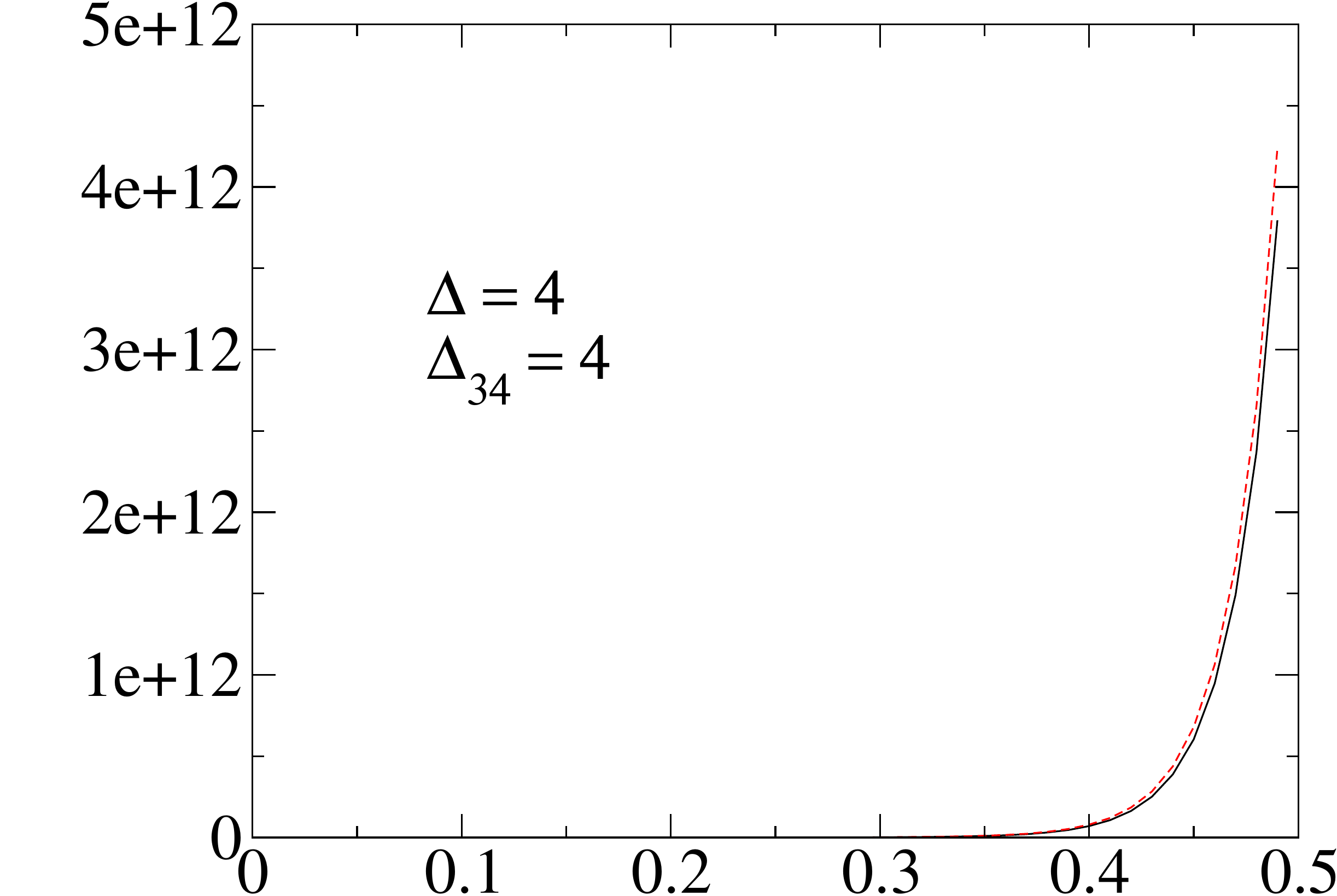} & \includegraphics[scale=0.2]{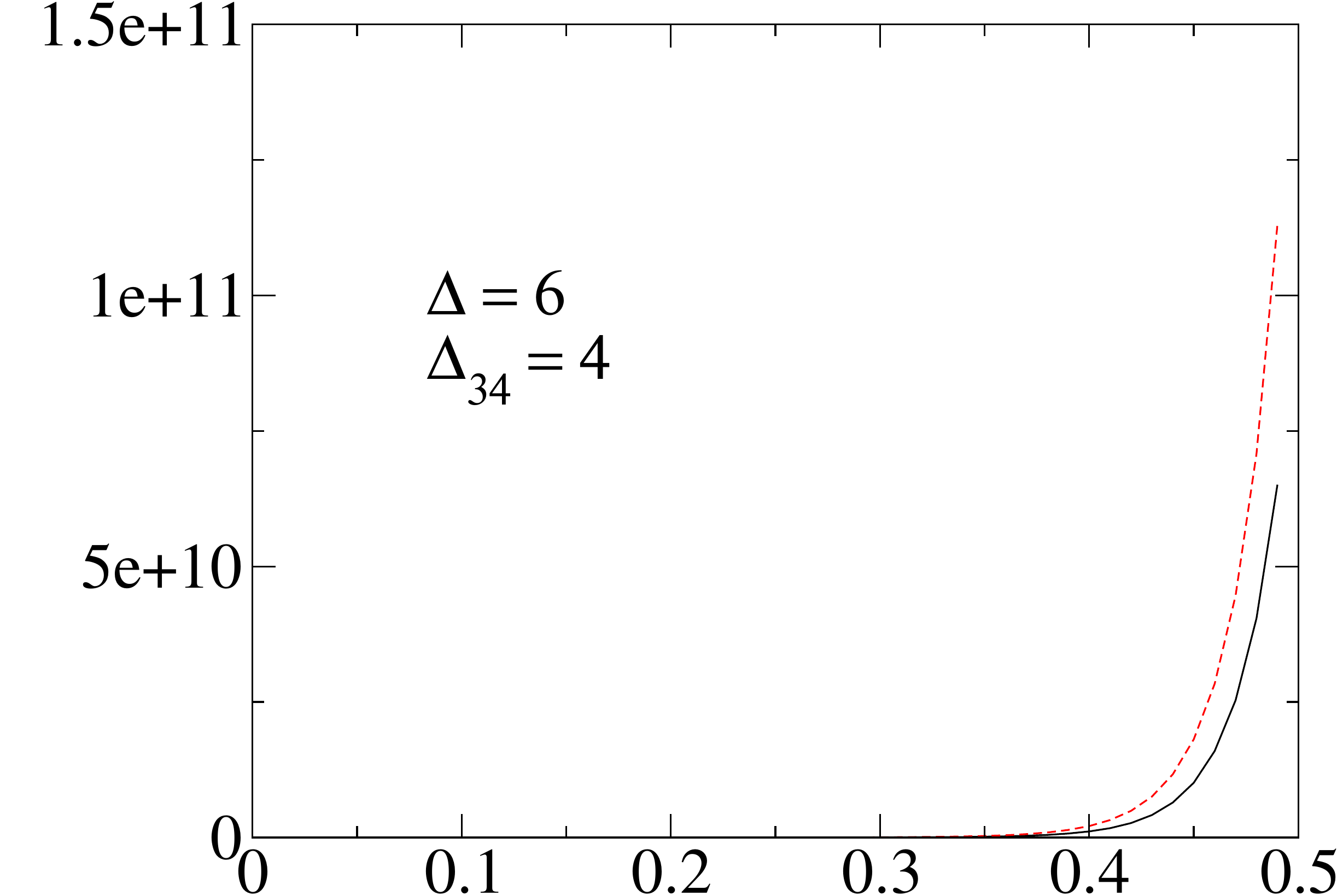} \\
\includegraphics[scale=0.2]{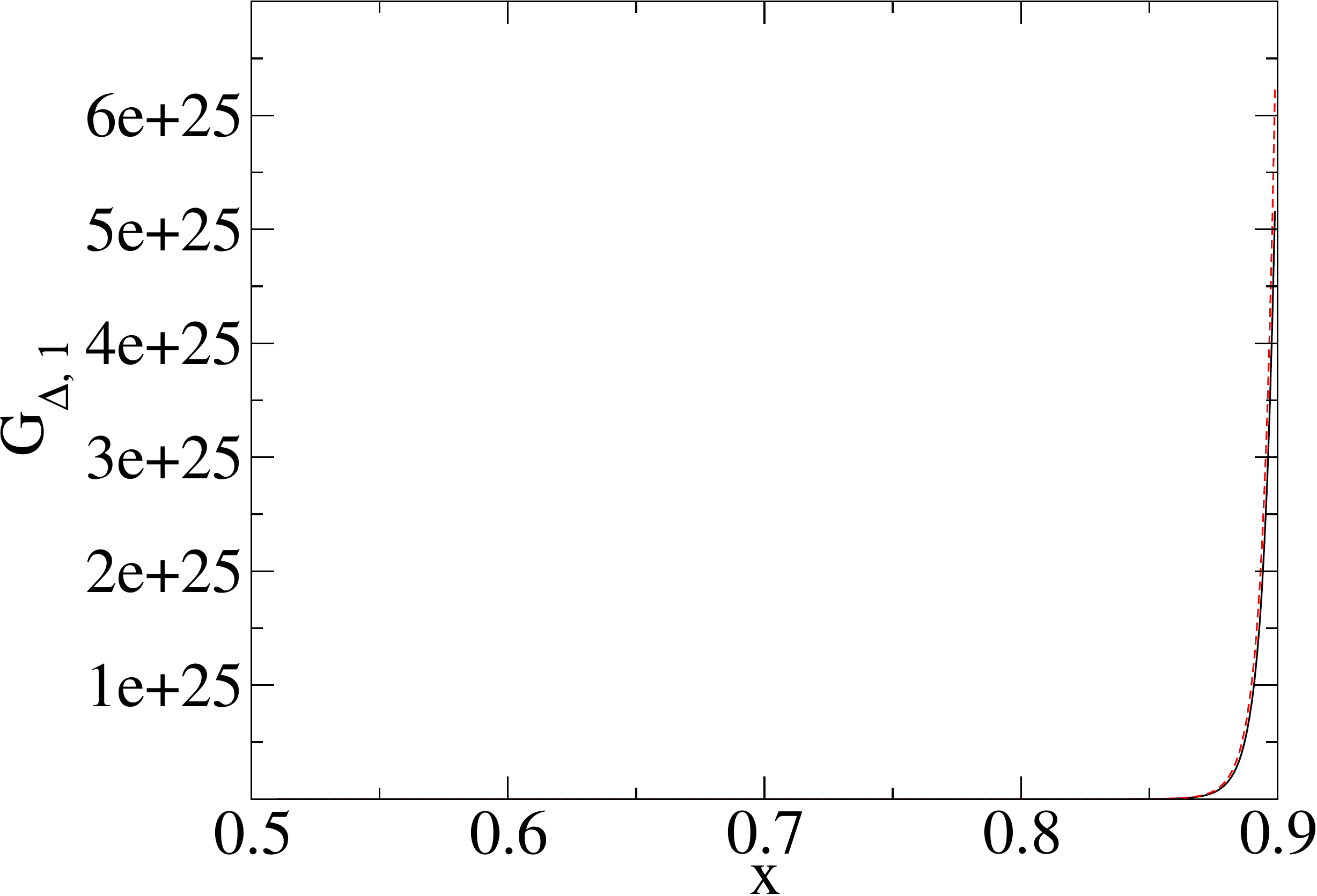} & \includegraphics[scale=0.2]{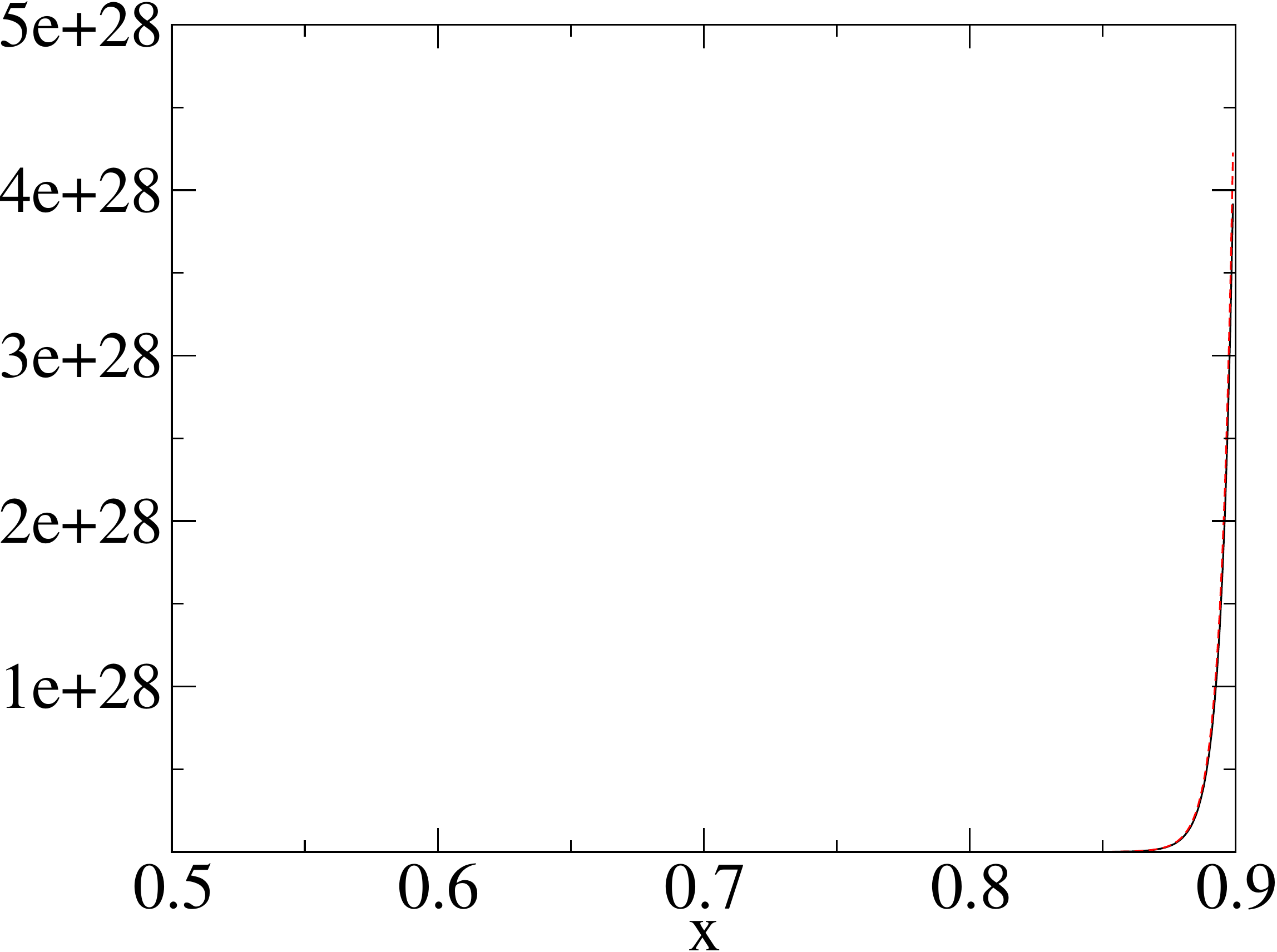} & \includegraphics[scale=0.2]{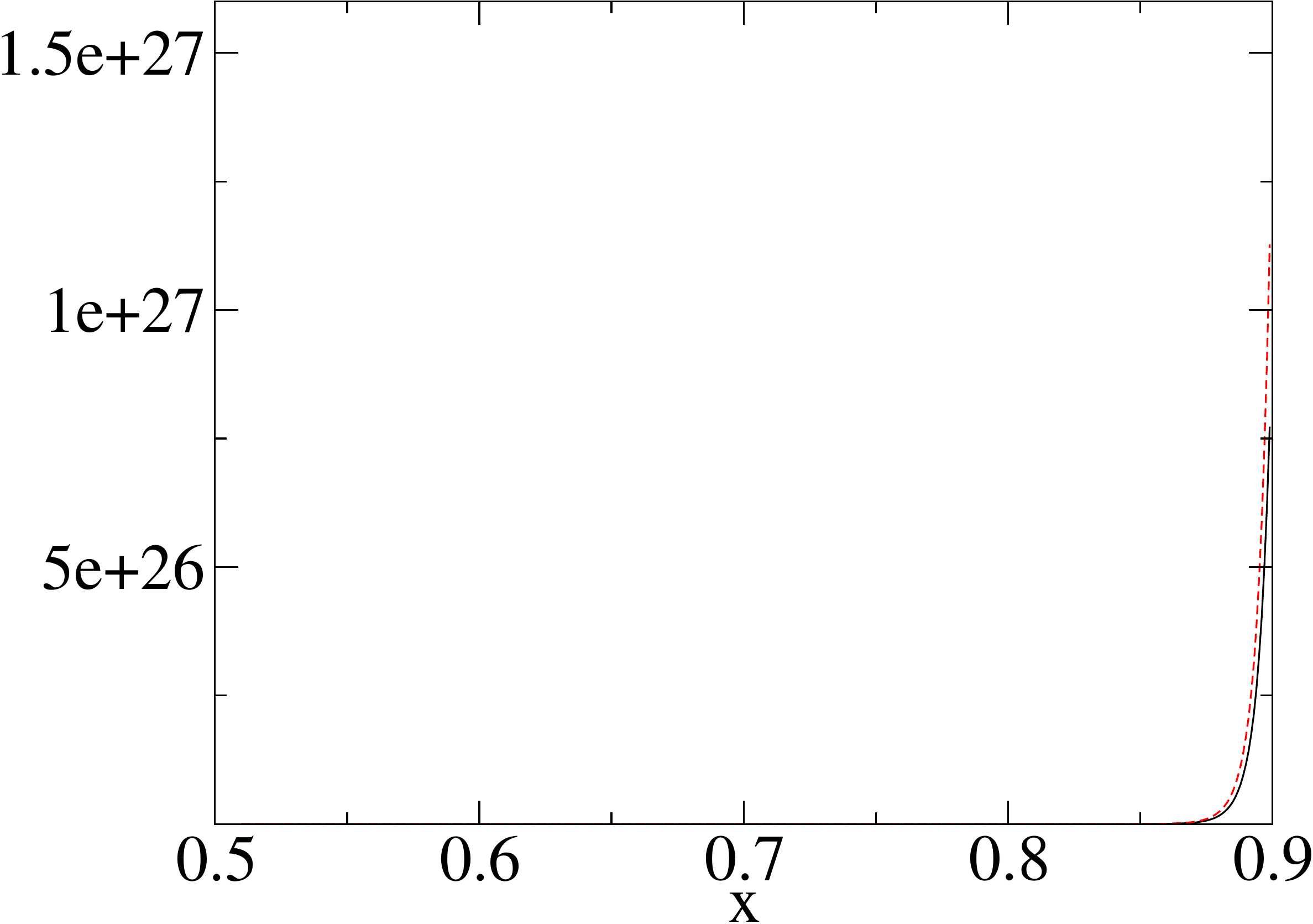}
\end{tabular}
\caption{For three different pairs $\left ( \Delta, \Delta_{34} \right )$, we compare the four dimensional conformal block (black line) to the asmyptotic version of it (red dotted line). The top plot has $x \in (0.1, 0.5)$, while the bottom plot has $x \in (0.5, 0.9)$. For each of these plots $z = \frac{1}{2}$, $\Delta_{12} = -40$ and $\ell = 1$.}
\label{plots}
\end{figure}
With (\ref{2f1-approx2}) in hand, we may expand the $d = 4$ solution
\begin{align}
G_{\mathcal{O}}(x, z) = \frac{ x^{\lambda_1 + 1}z^{\lambda_2} {}_2F_1 (\lambda_1 + a, \lambda_1 + b; 2\lambda_1; x) {}_2F_1 (\lambda_2 - 1 + a, \lambda_2 - 1 + b; 2\lambda_2 - 2; z) - (x \leftrightarrow z)}{x - z} \; . \label{4d-solution}
\end{align}
The $a^{2b - \lambda_1 - \lambda_2 + 1}$ term vanishes because the numerator is antisymmetric. We are left with:
\begin{equation}
G_{\mathcal{O}}(x, z) \sim (\lambda_1 - \lambda_2 + 1) (\lambda_1 + \lambda_2 - 2) \frac{\Gamma(2\lambda_1)\Gamma(2\lambda_2 - 2)}{\Gamma(\lambda_1 + b)\Gamma(\lambda_2 - 1 + b)} G^{(0)}_{\mathcal{O}}(x, z) \; . \label{result4d}
\end{equation}
The four dimensional block and its large $-\Delta_{12}$ limit are plotted in the Figure \ref{plots}. The $d = 6$ solution is
\begin{equation}
G_{\mathcal{O}}(x, z) = \sum_{p, q} a_{p, q} \frac{F_{p,q}^{-,3}(x, z)}{(x - z)^3} \label{6d-solution}
\end{equation}
with $a_{p, q}$ given by (\ref{exact-solutions}). Even though an individual $F_{p,q}^{-,3}$ is of order $a^{2b - \lambda_1 - \lambda_2 + 3}$, the coefficients conspire to cancel the first three asymptotic terms in the whole expression. To get $a^{2b - \lambda_1 - \lambda_2}$, we must expand each hypergeometric function in (\ref{6d-solution}) to fourth order. This will indeed give us the same form that the two dimensional block has after its hypergeometric functions are expanded to first order:
\begin{equation}
G_{\mathcal{O}}(x, z) \sim \frac{1}{6} (\lambda_1 - \lambda_2 + 2) (\lambda_1 - \lambda_2 + 3) (\lambda_1 + \lambda_2 - 3) (\lambda_1 + \lambda_2 - 4) \frac{\Gamma(2\lambda_1)\Gamma(2\lambda_2 - 4)}{\Gamma(\lambda_1 + b)\Gamma(\lambda_2 - 2 + b)} G^{(0)}_{\mathcal{O}}(x, z) \; . \label{result6d}
\end{equation}

\section{Solution for the coefficient}
From the Casimir differential equation, it would seem that the normalization for a conformal block is arbitrary. While performing the bootstrap however, it is essential for the blocks and the two point functions to be normalized in a consistent way. The convention that was first used in computing (\ref{2d-solution}) and (\ref{exact-solutions}) is
\begin{equation}
\lim_{x \rightarrow 0} \lim_{z \rightarrow 0} \frac{G_{\mathcal{O}}(x, z)}{x^{\lambda_1}z^{\lambda_2}} = 1 \; . \label{norm1}
\end{equation}
There is another common convention which is particularly natural for computing diagonal conformal blocks:
\begin{equation}
\lim_{z \rightarrow 0} \frac{G_{\mathcal{O}}(z, z)}{z^{\lambda_1 + \lambda_2}} = 1 \; . \label{norm2}
\end{equation}
These are not the same, as can be demonstrated by expanding the $d = 4$ block around $x = z$. Using the fact that hypergeometric functions tend to unity for small arguments,
\begin{eqnarray}
G_{\mathcal{O}}(x, z) &\approx& \frac{1}{x - z} \left [ x^{\lambda_1 + 1}z^{\lambda_2} - x^{\lambda_2}z^{\lambda_1 + 1} \right ] \nonumber \\
&=& \frac{1}{x - z} \left [ z^{\lambda_2} (z + (x - z))^{\lambda_1 + 1} - z^{\lambda_1 + 1} (z + (x - z))^{\lambda_2} \right ] \nonumber \\
&\approx& \frac{1}{x - z} \left [ z^{\lambda_1 + \lambda_2 + 1} \left ( 1 + (\lambda_1 + 1) \frac{x - z}{z} \right ) - z^{\lambda_1 + \lambda_2 + 1} \left ( 1 + \lambda_2 \frac{x - z}{z} \right ) \right ] \nonumber \\
&=& \frac{1}{x - z} \left [ (\lambda_1 - \lambda_2 + 1) z^{\lambda_1 + \lambda_2 + 1} \frac{x - z}{z} \right ] \nonumber \\
&=& (\lambda_1 - \lambda_2 + 1) z^{\lambda_1 + \lambda_2} \; . \label{norm-example}
\end{eqnarray}
We see that the four dimensional solution satisfying (\ref{norm1}) picks up a factor of $\lambda_1 - \lambda_2 + 1$ if we try to take the limit in (\ref{norm2}). This factor appears in our asymptotic (\ref{result4d}). Similarly, the six dimensional solution would pick up a factor of $\frac{1}{6} (\lambda_1 - \lambda_2 + 2) (\lambda_1 - \lambda_2 + 3)$ appearing in (\ref{result6d}) if we took the same limit. Notice that
\begin{eqnarray}
\lambda_1 - \lambda_2 + 1 &=& \frac{(2)_{\ell}}{(1)_{\ell}} \nonumber \\
\frac{1}{6} (\lambda_1 - \lambda_2 + 2) (\lambda_1 - \lambda_2 + 3) &=& \frac{(4)_{\ell}}{(2)_{\ell}} \; . \nonumber
\end{eqnarray}
In general, defining $\varepsilon = \frac{d - 2}{2}$, multiplication by $\frac{(2\varepsilon)_{\ell}}{(\varepsilon)_{\ell}}$ takes us from the normalization used in \cite{showk12a, hogervorst13a, hogervorst13b, dolan11} to the one used in \cite{dolan01, dolan04}.\footnote{Confusingly, the normalization $(-2)^{-\ell} \frac{(2\varepsilon)_{\ell}}{(\varepsilon)_{\ell}}$ appears in some parts of \cite{dolan01, dolan04} as well.} When solving for the constant $C^{(d)}_{\lambda_1, \lambda_2} \equiv C^{(d)}_{\mathcal{O}}$, we will switch to the (\ref{norm2}) normalization to avoid these extra factors.

\subsection{Base case}
Looking at (\ref{result2d}), (\ref{result4d}) and (\ref{result6d}), it seems reasonable to guess that
\begin{equation}
C^{(d)}_{\lambda_1, \lambda_2} = \frac{\Gamma(2\lambda_1)\Gamma(2\lambda_2 - 2\varepsilon)}{\Gamma(\lambda_1 + b)\Gamma(\lambda_2 - \varepsilon + b)} \frac{\Gamma(\lambda_1 + \lambda_2 - \varepsilon)}{\Gamma(\lambda_1 + \lambda_2 - 2\varepsilon)} \; . \label{coefficient}
\end{equation}
Taking the $z \rightarrow 0$ limit of (\ref{result3}) and trying to read off the coefficient would not be correct. This is because we have used an asymptotic expansion that only works away from zero. Going back to the saddle point approximation in (\ref{2f1-approx1}), the function we had to minimize is constant in this degenerate case. Proving our guess will therefore require more work.

To start, we will show that the conformal block for a spinless operator agrees with (\ref{coefficient}). Setting $\lambda_1 = \lambda_2 \equiv \lambda$, the double power series valid in any number of dimensions is:
\begin{align}
G_{\lambda, \lambda}(u, v) &= \sum_{m, n = 0}^{\infty} \frac{(\lambda - a)_m (\lambda - b)_m}{(2\lambda - \varepsilon)_m} \frac{(\lambda + a)_{m + n} (\lambda + b)_{m + n}}{(2\lambda)_{2m + n}} \frac{u^m}{m!} \frac{(1 - v)^n}{n!} \nonumber \\
&= \sum_{m = 0}^{\infty} \frac{(\lambda - a)_m (\lambda + a)_m (\lambda - b)_m (\lambda + b)_m}{(2\lambda - \varepsilon)_m (2\lambda)_{2m}} \frac{u^m}{m!} {}_2F_1(\lambda + a + m, \lambda + b + m; 2\lambda + 2m; 1 - v) \; . \nonumber
\end{align}
Expanding this hypergeometric function will give us a factor of $(\lambda + a + m)^{b - \lambda}$ which we may approximate as $a^{b - \lambda}$.
\begin{align}
G_{\lambda, \lambda}(u, v) &\sim \sum_{m = 0}^{\infty} \frac{(\lambda - a)_m (\lambda + a)_m (\lambda - b)_m (\lambda + b)_m}{(2\lambda - \varepsilon)_m (2\lambda)_{2m}} \frac{u^m}{m!} \frac{\Gamma(2\lambda + 2m)}{\Gamma(\lambda + b + m)} v^{-a - b} [(\lambda + a + m)(1 - v)]^{b - \lambda - m} \nonumber \\
&= \frac{\Gamma(2\lambda)}{\Gamma(\lambda + b)} u^{\lambda} v^{-a - b} (1 - v)^{b - \lambda} \sum_{m = 0}^{\infty} \frac{(\lambda - a)_m (\lambda + a)_m (\lambda - b)_m}{(2\lambda - \varepsilon)_m m!} \left ( \frac{u}{1 - v} \right )^m (\lambda + a + m)^{b - \lambda - m} \nonumber \\
&\sim \frac{\Gamma(2\lambda)}{\Gamma(\lambda + b)} a^{b - \lambda} u^{\lambda} v^{-a - b} (1 - v)^{b - \lambda} \sum_{m = 0}^{\infty} \frac{(\lambda - a)_m (\lambda + a)_m (\lambda - b)_m}{(\lambda + a + m)^m (2\lambda - \varepsilon)_m m!} \left ( \frac{u}{1 - v} \right )^m \nonumber
\end{align}
Since $(\lambda + a + m)^m$ and $(\lambda + a)_m$ are degree $m$ polynomials in $a$ with the same leading coefficient, we are justified in cancelling them when $a$ is large. This turns the sum into a hypergeometric function once again.
\begin{equation}
G_{\lambda, \lambda}(u, v) \sim \frac{\Gamma(2\lambda)}{\Gamma(\lambda + b)} a^{b - \lambda} u^{\lambda} v^{-a - b} (1 - v)^{b - \lambda} {}_2F_1 \left ( \lambda - a, \lambda - b; 2\lambda - \varepsilon; \frac{u}{1 - v} \right ) \nonumber
\end{equation}
Hypergeometric functions satisfy the identity ${}_2F_1(a, b; c; x) = (1 - x)^{-b} {}_2F_1 \left ( c - a, b; c; \frac{x}{x - 1} \right )$. We will use this twice so that the arguments have $a$ and $b$ appearing with positive signs:
\begin{align}
G_{\lambda, \lambda}(u, v) &\sim \frac{\Gamma(2\lambda)}{\Gamma(\lambda + b)} a^{b - \lambda} u^{\lambda} v^{-a - b} (1 - u - v)^{b - \lambda} \left ( \frac{1 - v}{1 - u - v} \right )^{\varepsilon - \lambda - a} \nonumber \\
& {}_2F_1 \left ( \lambda + a - \varepsilon, \lambda + b - \varepsilon; 2\lambda - \varepsilon; \frac{u}{1 - v} \right ) \nonumber \\
&\sim \frac{\Gamma(2\lambda)\Gamma(2\lambda - \varepsilon)}{\Gamma(\lambda + b)\Gamma(\lambda + b - \varepsilon)} a^{2b - 2\lambda} u^b v^{-a - b} \; . \label{spin0-step}
\end{align}
The expected coefficient thus appears for $\ell = 0$. A recursion will help us show that this continues to hold for higher spin.

\subsection{Inductive step}
Following \cite{dolan11}, three useful operators to define are:
\begin{eqnarray}
\mathcal{F}_0 &=& \frac{1}{x} + \frac{1}{z} - 1 \nonumber \\
\mathcal{F}_1 &=& (1 - x) \frac{\partial}{\partial x} + (1 - z) \frac{\partial}{\partial z} - a - b \nonumber \\
\mathcal{F}_2 &=& \frac{x - z}{xz} \left [ x^2 (1 - x) \frac{\partial^2}{\partial x^2} - x^2 \frac{\partial}{\partial x} - (x \leftrightarrow z)  \right ] \; . \label{f-operators}
\end{eqnarray}
These allow us to express the conformal block $G_{\lambda_1, \lambda_2}$ in terms of $G_{\lambda_1 \pm 1, \lambda_2 \pm 1}$. Recalling the definition of $B_p$ in (\ref{ab-def}), the recurrence relations also derived in \cite{dolan11} are:
\begin{align}
\mathcal{F}_0 G_{\lambda_1, \lambda_2} &= \frac{\lambda_1 - \lambda_2 + 2\varepsilon}{\lambda_1 - \lambda_2 + \varepsilon} G_{\lambda_1, \lambda_2 - 1} + \frac{\lambda_1 - \lambda_2}{\lambda_1 - \lambda_2 + \varepsilon} G_{\lambda_1 - 1, \lambda_2} + \frac{(\lambda_1 + \lambda_2 - 1)(\lambda_1 + \lambda_2 - 2\varepsilon)}{(\lambda_1 + \lambda_2 - 1 - \varepsilon)(\lambda_1 + \lambda_2 - \varepsilon)} \nonumber \\
& \left ( \frac{\lambda_1 - \lambda_2 + 2\varepsilon}{\lambda_1 - \lambda_2 + \varepsilon} B_{\lambda_1} G_{\lambda_1 + 1, \lambda_2} + \frac{\lambda_1 - \lambda_2}{\lambda_1 - \lambda_2 + \varepsilon} B_{\lambda_2 - \varepsilon} G_{\lambda_1, \lambda_2 + 1} \right ) \nonumber \\
& - \frac{ab}{2} \frac{\lambda_1 (\lambda_1 - 1) + \lambda_2 (\lambda_2 - 1 - 2\varepsilon) + 2\varepsilon}{\lambda_1 (\lambda_1 - 1)(\lambda_2 - \varepsilon)(\lambda_2 - 1 - \varepsilon)} G_{\lambda_1, \lambda_2} \label{recursion3a} \\
\mathcal{F}_1 G_{\lambda_1, \lambda_2} &= \lambda_2 \frac{\lambda_1 - \lambda_2 + 2\varepsilon}{\lambda_1 - \lambda_2 + \varepsilon} G_{\lambda_1, \lambda_2 - 1} + (\lambda_1 + \varepsilon) \frac{\lambda_1 - \lambda_2}{\lambda_1 - \lambda_2 + \varepsilon} G_{\lambda_1 - 1, \lambda_2} - \frac{(\lambda_1 + \lambda_2 - 1)(\lambda_1 + \lambda_2 - 2\varepsilon)}{(\lambda_1 + \lambda_2 - 1 - \varepsilon)(\lambda_1 + \lambda_2 - \varepsilon)} \nonumber \\
& \left ( (\lambda_1 - 1 - \varepsilon) \frac{\lambda_1 - \lambda_2 + 2\varepsilon}{\lambda_1 - \lambda_2 + \varepsilon} B_{\lambda_1} G_{\lambda_1 + 1, \lambda_2} + (\lambda_2 - 1 - 2\varepsilon) \frac{\lambda_1 - \lambda_2}{\lambda_1 - \lambda_2 + \varepsilon} B_{\lambda_2 - \varepsilon} G_{\lambda_1, \lambda_2 + 1} \right ) \nonumber \\
& - (1 + \varepsilon) \frac{ab}{2} \frac{\lambda_1 (\lambda_1 - 1) + \lambda_2 (\lambda_2 - 1 - 2\varepsilon) + 2\varepsilon}{\lambda_1 (\lambda_1 - 1)(\lambda_2 - \varepsilon)(\lambda_2 - 1 - \varepsilon)} G_{\lambda_1, \lambda_2} \label{recursion3b} \\
\mathcal{F}_2 G_{\lambda_1, \lambda_2} &= (\lambda_1 + \lambda_2 - 1) \frac{(\lambda_1 - \lambda_2)(\lambda_1 - \lambda_2 + 2\varepsilon)}{\lambda_1 - \lambda_2 + \varepsilon} \bigg [ G_{\lambda_1, \lambda_2 - 1} - G_{\lambda_1 - 1, \lambda_2} - \bigg. \nonumber \\
& \bigg. \frac{(\lambda_1 + \lambda_2 - 2\varepsilon)(\lambda_1 + \lambda_2 - 1 - 2\varepsilon)}{(\lambda_1 + \lambda_2 - \varepsilon)(\lambda_1 + \lambda_2 - 1 - \varepsilon)} \left ( B_{\lambda_1} G_{\lambda_1 + 1, \lambda_2} - B_{\lambda_2 - \varepsilon} G_{\lambda_1, \lambda_2 + 1} \right ) \bigg ] \nonumber \\
& - \frac{ab}{2} \frac{(\lambda_1 - \lambda_2)(\lambda_1 - \lambda_2 + 2\varepsilon)(\lambda_1 + \lambda_2 - 1)(\lambda_1 + \lambda_2 - 1 - 2\varepsilon)}{\lambda_1 (\lambda_1 - 1)(\lambda_2 - \varepsilon)(\lambda_2 - 1 - \varepsilon)} G_{\lambda_1, \lambda_2} \; . \label{recursion3c}
\end{align}
After shifting $\lambda_2 \mapsto \lambda_2 + 1$, any two of these relations can be combined to eliminate $G_{\lambda_1 + 1, \lambda_2 + 1}$. The following equations come from combining (\ref{recursion3a}) with (\ref{recursion3c}) and (\ref{recursion3b}) respectively:
% Also hackish.
\begin{align}
& \frac{(\lambda_1 + \lambda_2 - \varepsilon)(\lambda_1 - \lambda_2 - 1 + 2\varepsilon)}{\lambda_1 - \lambda_2 - 1 + \varepsilon} G_{\lambda_1, \lambda_2} = \frac{\varepsilon(2\lambda_1 - 1)}{\lambda_1 - \lambda_2 - 1 + \varepsilon} G_{\lambda_1 - 1, \lambda_2 + 1} \nonumber \\
& \;\;\;\;\;\;\;\;\;\;\; + \frac{1}{2} \left [ (\lambda_1 + \lambda_2 - 2\varepsilon) \left ( \mathcal{F}_0 + \frac{ab}{2} \frac{\lambda_1 (\lambda_1 - 1) + (\lambda_2 + 1) (\lambda_2 - 2\varepsilon) + 2\varepsilon}{\lambda_1 (\lambda_1 - 1)(\lambda_2 + 1 - \varepsilon)(\lambda_2 - \varepsilon)} \right ) \right. \nonumber \\
& \;\;\;\;\;\;\;\;\;\;\; \left. + \frac{1}{\lambda_1 - \lambda_2 - 1} \left ( \mathcal{F}_2 + \frac{ab}{2} \frac{(\lambda_1 - \lambda_2 - 1)(\lambda_1 - \lambda_2 - 1 + 2\varepsilon)(\lambda_1 + \lambda_2)(\lambda_1 + \lambda_2 - 2\varepsilon)}{\lambda_1 (\lambda_1 - 1)(\lambda_2 + 1 - \varepsilon)(\lambda_2 - \varepsilon)} \right ) \right ] G_{\lambda_1, \lambda_2 + 1} \nonumber \\
& \;\;\;\;\;\;\;\;\;\;\; - \frac{(\lambda_1 + \lambda_2)(\lambda_1 + \lambda_2 - 2\varepsilon)(\lambda_1 + \lambda_2 + 1 - 2\varepsilon)}{(\lambda_1 + \lambda_2 - \varepsilon)(\lambda_1 + \lambda_2 + 1 - \varepsilon)} B_{\lambda_2 + 1 - \varepsilon} G_{\lambda_1, \lambda_2 + 2} \label{recursion4a} \\
& \frac{(\lambda_1 + \lambda_2 - \varepsilon)(\lambda_1 - \lambda_2 - 1 + 2\varepsilon)}{\lambda_1 - \lambda_2 - 1 + \varepsilon} G_{\lambda_1, \lambda_2} = \bigg [ (\lambda_1 - 1 - \varepsilon) \mathcal{F}_0 + \mathcal{F}_1 \bigg. \nonumber \\
& \;\;\;\;\;\;\;\;\;\;\; \bigg. + \frac{ab}{2} \frac{\lambda_1 (\lambda_1 - 1) + (\lambda_2 + 1) (\lambda_2 - 2\varepsilon) + 2\varepsilon}{(\lambda_1 - 1)(\lambda_2 + 1 - \varepsilon)(\lambda_2 - \varepsilon)} \bigg ] G_{\lambda_1, \lambda_2 + 1} - (\lambda_1 - \lambda_2 - 1) \nonumber \\
& \;\;\;\;\;\;\;\;\;\;\; \left [ \frac{(\lambda_1 + \lambda_2)(\lambda_1 + \lambda_2 + 1 - 2\varepsilon)}{(\lambda_1 + \lambda_2 - \varepsilon)(\lambda_1 + \lambda_2 + 1 - \varepsilon)} B_{\lambda_2 + 1 - \varepsilon} G_{\lambda_1, \lambda_2 + 2} + \frac{2\lambda_1 - 1}{\lambda_1 - \lambda_2 - 1 + \varepsilon} G_{\lambda_1 - 1, \lambda_2 + 1} \right ] \; . \label{recursion4b}
\end{align}
In the appendix of \cite{showk12a}, these same recurrence relations appear for the case $a = b = 0$. They give spin $\ell$ operators on the left in terms of spin $\ell - 1$ and $\ell - 2$ operators on the right. Notice that the last term in (\ref{recursion4b}) vanishes when $\lambda_1 - \lambda_2 = 1$. Therefore, this equation can be used to convert the spin 0 block $G_{\lambda + \frac{1}{2}, \lambda + \frac{1}{2}}$ into the spin 1 block $G_{\lambda + \frac{1}{2}, \lambda - \frac{1}{2}}$. Using this to find the large $a$ spin 1 block from (\ref{spin0-step}),
\begin{eqnarray}
G_{\lambda + \frac{1}{2}, \lambda - \frac{1}{2}} &=& \frac{1}{4\lambda - 2\varepsilon} \left [ \left ( \lambda - \frac{1}{2} - \varepsilon \right ) \mathcal{F}_0 + \mathcal{F}_1 + \frac{ab}{2} \frac{\left ( \lambda + \frac{1}{2} \right ) (2\lambda - 1 - 2\varepsilon) + 2\varepsilon}{\left ( \lambda - \frac{1}{2} \right ) \left ( \lambda + \frac{1}{2} - \varepsilon \right ) \left ( \lambda - \frac{1}{2} - \varepsilon \right )} \right ] G_{\lambda + \frac{1}{2}, \lambda + \frac{1}{2}} \nonumber \\
&\sim& \frac{1}{4\lambda - 2\varepsilon} \left [ (1 - x)\frac{\partial}{\partial x} + (1 - z)\frac{\partial}{\partial z} - a + \frac{ab}{2} \frac{\left ( \lambda + \frac{1}{2} \right ) (2\lambda - 1 - 2\varepsilon) + 2\varepsilon}{\left ( \lambda - \frac{1}{2} \right ) \left ( \lambda + \frac{1}{2} - \varepsilon \right ) \left ( \lambda - \frac{1}{2} - \varepsilon \right )} \right ] G_{\lambda + \frac{1}{2}, \lambda + \frac{1}{2}} \nonumber \\
&\sim& \frac{1}{4\lambda - 2\varepsilon} \left [ a + \frac{ab}{2} \frac{\left ( \lambda + \frac{1}{2} \right ) (2\lambda - 1 - 2\varepsilon) + 2\varepsilon}{\left ( \lambda - \frac{1}{2} \right ) \left ( \lambda + \frac{1}{2} - \varepsilon \right ) \left ( \lambda - \frac{1}{2} - \varepsilon \right )} \right ] G_{\lambda + \frac{1}{2}, \lambda + \frac{1}{2}} \nonumber \\
&=& \frac{a}{4\lambda - 2\varepsilon} \frac{\lambda - \frac{1}{2} + b - \varepsilon}{\lambda - \frac{1}{2} - \varepsilon} G_{\lambda + \frac{1}{2}, \lambda + \frac{1}{2}} \; . \nonumber
\end{eqnarray}
Inserting the $u$ and $v$ dependence,
\begin{eqnarray}
G_{\lambda + \frac{1}{2}, \lambda - \frac{1}{2}} (u, v) &\sim& \frac{\lambda - \frac{1}{2} + b - \varepsilon}{(2\lambda - \varepsilon)(2\lambda - 1 - 2\varepsilon)} \frac{\Gamma(2\lambda + 1)\Gamma(2\lambda + 1 - \varepsilon)}{\Gamma \left ( \lambda + \frac{1}{2} + b \right ) \Gamma \left ( \lambda + \frac{1}{2} + b - \varepsilon \right )} a^{2b - 2\lambda} u^b v^{-a - b} \nonumber \\
&=& \frac{1}{2\lambda - 1 - 2\varepsilon} \frac{\Gamma(2\lambda + 1)\Gamma(2\lambda - \varepsilon)}{\Gamma \left ( \lambda + \frac{1}{2} + b \right ) \Gamma \left ( \lambda - \frac{1}{2} + b - \varepsilon \right )} a^{2b - 2\lambda} u^b v^{-a - b} \; . \label{spin1-step}
\end{eqnarray}
Since the coefficient (\ref{coefficient}) holds for $\ell = 0$ and $\ell = 1$, we may use (\ref{recursion4a}) to show that it holds for all $\ell$. Instead of working with this equation for general $x$ and $z$, it is sufficient to restrict it to the diagonal $x = z$. The advantage of this is that $\mathcal{F}_2$ drops out giving us a recursion with no derivatives. This diagonal limit allowed \cite{showk12a} to solve the recursion with $a = b = 0$. A solution for $a = 0$ but $b \neq 0$ was later found in \cite{hogervorst13b}. We will instead drop terms surpressed by powers of $a$ so that none of the parameters have to be zero. This means that $G_{\lambda_1, \lambda_2 + 1}$ only survives when multiplied by $a$ and $G_{\lambda_1, \lambda_2 + 2}$ only survives when multiplied by $a^2$. This yields
\begin{align}
& \frac{(\lambda_1 + \lambda_2 - \varepsilon)(\lambda_1 - \lambda_2 - 1 + 2\varepsilon)}{\lambda_1 - \lambda_2 - 1 + \varepsilon} G^{(0)}_{\lambda_1, \lambda_2}(z, z) = \frac{\varepsilon(2\lambda_1 - 1)}{\lambda_1 - \lambda_2 - 1 + \varepsilon} G^{(0)}_{\lambda_1 - 1, \lambda_2 + 1}(z, z) \nonumber \\
& \;\;\;\;\;\;\;\;\;\;\; + \frac{ab (\lambda_1 + \lambda_2 - 2\varepsilon) [ \lambda_1 (\lambda_1 - 1) + \lambda_2 (\lambda_2 + 1 - 2\varepsilon) + (\lambda_1 - \lambda_2 - 1 + 2\varepsilon)(\lambda_1 + \lambda_2) ]}{4\lambda_1 (\lambda_1 - 1)(\lambda_2 + 1 - \varepsilon)(\lambda_2 - \varepsilon)} G^{(0)}_{\lambda_1, \lambda_2 + 1}(z, z) \nonumber \\
& \;\;\;\;\;\;\;\;\;\;\; + \frac{a^2 (\lambda_1 + \lambda_2)(\lambda_1 + \lambda_2 - 2\varepsilon)(\lambda_1 + \lambda_2 + 1 - 2\varepsilon) [ (\lambda_2 + 1 - \varepsilon)^2 - b^2 ]}{4 (\lambda_1 + \lambda_2 - \varepsilon)(\lambda_1 + \lambda_2 + 1 - \varepsilon) (\lambda_2 + 1 - \varepsilon)^2(2\lambda_2 + 3 - 2\varepsilon)(2\lambda_2 + 1 - 2\varepsilon)} G^{(0)}_{\lambda_1, \lambda_2 + 2}(z, z) \; . \label{recursion5}
\end{align}
Cancelling the factors of $a^{2b - \lambda_1 - \lambda_2} z^{2b} (1 - z)^{-2a - 2b}$ on either side, this becomes a recurrence relation for the coefficients:
\begin{align}
& \frac{(\lambda_1 + \lambda_2 - \varepsilon)(\lambda_1 - \lambda_2 - 1 + 2\varepsilon)}{\lambda_1 - \lambda_2 - 1 + \varepsilon} C^{(d)}_{\lambda_1, \lambda_2} = \frac{\varepsilon(2\lambda_1 - 1)}{\lambda_1 - \lambda_2 - 1 + \varepsilon} C^{(d)}_{\lambda_1 - 1, \lambda_2 + 1} \nonumber \\
& \;\;\;\;\;\;\;\;\;\;\; + \frac{b (\lambda_1 + \lambda_2 - 2\varepsilon) [ \lambda_1 (\lambda_1 - 1) + \lambda_2 (\lambda_2 + 1 - 2\varepsilon) + (\lambda_1 - \lambda_2 - 1 + 2\varepsilon)(\lambda_1 + \lambda_2) ]}{4\lambda_1 (\lambda_1 - 1)(\lambda_2 + 1 - \varepsilon)(\lambda_2 - \varepsilon)} C^{(d)}_{\lambda_1, \lambda_2 + 1} \nonumber \\
& \;\;\;\;\;\;\;\;\;\;\; + \frac{(\lambda_1 + \lambda_2)(\lambda_1 + \lambda_2 - 2\varepsilon)(\lambda_1 + \lambda_2 + 1 - 2\varepsilon) [ (\lambda_2 + 1 - \varepsilon)^2 - b^2 ]}{4 (\lambda_1 + \lambda_2 - \varepsilon)(\lambda_1 + \lambda_2 + 1 - \varepsilon) (\lambda_2 + 1 - \varepsilon)^2(2\lambda_2 + 3 - 2\varepsilon)(2\lambda_2 + 1 - 2\varepsilon)} C^{(d)}_{\lambda_1, \lambda_2 + 2} \; . \label{recursion6}
\end{align}
It is not difficult to check that (\ref{recursion6}) is solved by (\ref{coefficient}). It follows that (\ref{coefficient}) gives the correct coefficient for any spin.

\section{Discussion and summary}
Using an asymptotic expansion in the large parameter $a = -\frac{1}{2}\Delta_{12}$, we have managed to rewrite the complicated function $G_{\mathcal{O}}(u, v)$ as $G^{(0)}_{\mathcal{O}}(u, v)$ which is monomial in $u$ and $v$. Specifically, it is (\ref{result3}) with a prefactor given by (\ref{coefficient}). We now return to the question of whether the conformal bootstrap becomes more tractable in this large parameter case.

Most studies involving the conformal bootstrap have considered four point functions $\left < \phi_1(x_1) \phi_2(x_2) \phi_3(x_3) \phi_4(x_4) \right >$ where the fields have equal or comparable scaling dimensions. Such a correlator will only produce large dimension differences when the exchanged primary $\mathcal{O}$ is a heavy operator. The infinite set of constraints from crossing symmetry requires one to discard all sufficiently heavy operators from the linear programming problem before it can be solved numerically. For example, \cite{rattazzi08} considered scaling dimensions of at most 20 with the hope that the rest have a negligible contribution. This was made rigorous in \cite{pappadopulo12} which found that for fixed $\Delta_1$ and $\Delta_2$, the OPE coefficient $\lambda_{12\mathcal{O}}$ decreases rapidly as $\Delta \rightarrow \infty$.

In the limit we have taken, the largest $\lambda_{12\mathcal{O}}$ coefficient in the conformal block expansion will be smaller than any that were kept by previous studies. These coefficients will again need to be cut off once they reach an even smaller size. A key question is whether this can be done before the dimension of $\mathcal{O}$ approaches $\left | \Delta_{12} \right |$. This would enable us to use
\begin{equation}
\sum_{\mathcal{O}} \lambda_{12\mathcal{O}} \lambda_{34\mathcal{O}} G^{(0)}_{\mathcal{O}}(u, v) \label{approximate-sum}
\end{equation}
to approximate the unknown part of the four point functon $G(u, v)$. The problem is that the $\Delta$ appearing in this sum will eventually become large enough to invalidate (\ref{result3}). With the tail of the series not controlled by our approximation, one has to worry about whether it still converges.

Powers of $u$ and $v$ in (\ref{result3}) multiply $a^{2b - \Delta}$ which decreases exponentially with $\Delta$. However, for fixed $\ell$, $b$ and $d$, (\ref{coefficient}) is asymptotic to $\left ( \frac{\Gamma(\Delta)}{\Gamma \left ( \frac{\Delta}{2} \right )} \right )^2$ which has a faster than exponential increase. This would have to be cancelled by the decrease of $\lambda_{12\mathcal{O}} \lambda_{34\mathcal{O}}$ for the terms in (\ref{approximate-sum}) to go to zero. As argued by \cite{pappadopulo12}, in a basis for the CFT Hilbert space, the number of local operators with dimension $\Delta$ is exponential in $\Delta^{\frac{d - 1}{d}}$. When these same states are weighted by their OPE coefficients, they instead find a power law, suggesting that the rate $\lambda_{12\mathcal{O}} \lambda_{34\mathcal{O}} \rightarrow 0$ cannot be faster than exponential. We are therefore using an approximation that changes the convergence properties of the conformal block expansion. This makes it doubtful that it is valid to truncate the series in a regime where the approximation still holds.

Even if we are not able to take the large $a$ limit of all the conformal blocks in $G(u, v)$, it was emphasized in \cite{fitzpatrick12a} that a single conformal block bears some relation to an AdS Feynman diagram or Witten diagram. The idea is to find the poles $\delta_m$ that the conformal blocks have in Mellin space and write
\begin{equation}
\sum_m \mathrm{Res} \left ( W_{\mathcal{O}}, \delta_m \right ) \frac{1}{\delta - \delta_m} \; . \label{pole-sum}
\end{equation}
The difference between this sum over poles and a corresponding Witten diagram would not be obvious in position space. An advantage of Mellin space is that diagrams like the ones in \cite{penedones10} will only differ from (\ref{pole-sum}) by polynomials.

Up to a normalization, a conformal partial wave in Mellin space is given by:
\begin{equation}
W_{\mathcal{O}}(\delta_{ij}) = e^{\pi i \left ( \frac{d}{2} - \Delta \right )} \left ( e^{\pi i (\delta + \Delta - d)} - 1 \right ) \frac{\Gamma \left ( \frac{\Delta - \ell - \delta}{2} \right ) \Gamma \left ( \frac{d - \Delta - \ell - \delta}{2} \right )}{\Gamma \left ( \frac{\Delta_1 + \Delta_2 - \delta}{2} \right ) \Gamma \left ( \frac{\Delta_3 + \Delta_4 - \delta}{2} \right )} P^{(d)}_{\ell}(\delta_{ij}) \; . \label{mellin-block}
\end{equation}
Note that dimension differences are encoded in the variables through $\sum_{i \neq j} \delta_{ij} = \Delta_j$. While this formula makes the locations of the poles explicit in arbitrary dimension, their residues depend crucially on the Mack polynomials $P^{(d)}_{\ell}$ which are defined by rather complicated recurrence relations. One could probably gain more information about what they are for large $a$ by inverting the Mellin-Barnes integral representation of our result:
\begin{equation}
W_{\mathcal{O}}(P_1, P_2, P_3, P_4) = \int_{-i\infty}^{i\infty} W_{\mathcal{O}}(\delta_{ij}) \prod_{i < j} \Gamma (\delta_{ij}) P_{ij}^{-\delta_{ij}} \textup{d}\delta_{ij} \; . \label{mellin-transform}
\end{equation}

The presence of heavy operators makes the conformal block we have considered very different from the ones related to scattering processes in \cite{fitzpatrick12a}. This is because correlation functions involving heavy operators are not described by Witten diagrams. From the generating functional for CFT correlators, $\exp \left ( \int_{\mathbb{R}^d} \mathcal{O}_i(x) \phi_{i, \partial}(x) \textup{d}x \right )$, it is clear that the dimensions of a bulk field and its dual operator must add up to $d$. If $\Delta_i$ is large but $d$ is not, the dual field is not relevant in a low energy effective action for the bulk.

A case that falls under our limit is $\left < \phi(x_1) \Phi(x_2) \Phi(x_3) \Phi(x_4) \right >$ where $\Delta_{\Phi} \gg \Delta_{\phi}$. If the boundary CFT is in a state corresponding to these operators, its gravity dual is essentially a test particle travelling through a warped geometry. This should only be asymptotically AdS because of the backreaction caused by the three heavy operators. If the connection between conformal blocks and scattering continues to hold in this case, the pole structure of our result can be used to probe the geometries associated with certain heavy operators in the CFT.

\section*{Acknowledgements}
This work would not have been possible without Mark Van Raamsdonk who first introduced me to the conformal bootstrap. I would also like to thank the participants of the ``Modern Developments in M-Theory'' workshop, particularly Neil Lambert, Leonardo Rastelli, Daniel Robbins and James Sully, for useful discussions about conformal field theory. Leonardo Rastelli and Mark Van Raamsdonk gave feedback on the draft.

\bibliographystyle{unsrt}
\bibliography{references}

\end{document}